\documentstyle[12pt]{article}                       

\begin{document}

\hfill{file : forarxiv[sctheory.tex]}

\vspace{1cm}

\centerline{\bf  Basic Foundations of the Microscopic Theory of
Superconductivity }

\vspace{0.7cm}

\centerline{\bf Yatendra S. Jain}

\bigskip
\centerline{Department of Physics,}

\centerline{North-Eastern Hill University, Shillong - 793 022, India}

\vspace{0.7cm}

\begin{abstract}
A new approach based on macro-orbital representation of a
conduction electron in a solid has been used to discover
some untouched aspects of the phonon induced attraction
between two electrons and to lay the basic foundations of
a general theory of superconductivity applicable to widely
different solids.  To this effect we first analyze the net
hamiltonian, $H(N)$, of $N$ conduction electrons to identify
its universal part, $H_o(N)$ ({\it -independent of the
nature of a specific solid or a specific class of solids}),
and then study the states of $H_o(N)$ to conclude that
superconductivity originates, basically, from an inter-play
between the zero-point force ($f_o$) of conduction electrons
in their ground state and the inter-atomic forces ($f_a$)
which decide the lattice structure.  This renders a kind
of mechanical strain in the lattice which serves as the main
source of phonon induced inter-electron attraction responsible
for the formation of Cooper type pairs and the onset of
superconductivity below certain temperature $T_c$.  We
determine the binding energy of such pairs and find a relation
for $T_c$ which not only accounts for the highest experimental
$T_c \approx 135$K that we know to-day but also indicates that
superconductivity may, in principle, occur at room temperature.
It is evident that electrical strain in the lattice
({\it i.e.} electrical polarization of the lattice
constituents produced by the charge of conducting electrons)
can have an added contribution to the phonon induced attraction
of two electrons.  Our theoretical framework not only
incorporates BCS model but also provides microscopic basis for
the two well known phenomenologies of superconductivity,
{\it viz.}, the two fluid theory and $\Psi-$theory.  In
addition, it also corroborates a recent idea that
superconducting transition is basically a quantum phase
transition.

\end{abstract}



\bigskip
\centerline{\bf 1.0 Introduction}

\bigskip
The experimental discovery of high $T_c$ (HTC) superconductivity 
in 1987 [1] came as a great surprise to the physics community,
basically, for its challenge to the Bardeen, Cooper and Schrieffer
(BCS) model [2] which had emerged as a highly successful theory of
superconductors that we knew at that time.  Consequently, HTC
systems became a subject of intense research activity and
thousands of experimental and theoretical papers have been
published over the last eighteen years.  While the important
results of various experimental studies on HTC systems are reviewed
in [3-13], the status of our present theoretical understanding is
elegantly summed up in [3, 12-20]; references to other reviews and
important research articles can be obtained from [3-21]. 

\bigskip
Several theoretical models based on widely different exotic
ideas have been worked out, since no single mechanism could be
identified as the basic origin of different properties of HTC
systems.  We have theories based on Hubbard model [17, 22-24] ,
spin bag theories [25, 26], antiferromagnetic Fermi-liquid theory
[27], $d_{x^2-y^2}$ theories [28, 29], anyon theory [30],
bipolaron theory [31] and theories based on the proximity effects
of quantum phase transition [15] and it may be mentioned that
this list is not exhaustive; the references to many other models
can be traced from [3, 12-21].

\bigskip
It is evident that, even after a period of nearly two decades
of the discovery of HTC systems, the goal of having a single
microscopic theory of superconductivity is far from being
achieved.  Incidentally, the process of achieving this goal has
been frustrated further by certain experimental results,
{\it viz.} : (i) the coexistence of superconductivity with
ferro-magnetism [32], (ii) superconductivity of $MgB_2$ at $T_c
(\approx 39)$K [33], (iii) pressure/strain induced
superconductivity [34], (iv) stripes of charges in a HTC system
[35], enhancement of superconductivity by nano-engineered
magnetic field in the form of tiny magnetic dots [36], {\it etc.}
as well as by interesting theoretical models which consider two
energy gaps [37], formation of Cooper type pairs through
spin-spin interaction [38], triplet p-wave pairing and singlet
d-wave pairing [39], {\it etc.} for specific superconductors.
As such we either have a system specific theory or a class
({\it i.e.} a set of superconducting solids) specific theory
of superconductivity and numerous ideas that have greatly
muddled the selection of right idea(s) which may help in
developing a unified single microscopic theory ({\it preferably
incorporating BCS theory}) of the phenomenon.  However, we found
a way-out by using a non-conventional approach to the problem
based on the macro-orbital representation of a quantum
particle [40].   Our approach makes no assumption about the
nature of the order parameter of superconducting transition
or the nature and strength of the interaction responsible
for the formation of Cooper type pairs.  It simply banks
upon the solutions of the Schr\"{o}dinger equation of
$N$ conduction electrons.

\bigskip
We note that conduction electrons form a kind of Fermi fluid
which flows through the lattice structure of a solid.  To a good
approximation, each electron can be identified as a freely moving
particle which can be represented by a plane wave unless it
suffers collisions with other electron(s) or lattice.  It is
argued that electrostatic screening effect of the lattice
significantly reduces the strength and range of electron-electron
repulsion and thereby facilitates the formation of Cooper pairs
basically responsible for the phenomenon [2].  However, these
effects in certain superconductors ({\it e.g.} in HTC systems) are
found to be relatively weak and each theoretical model of such
superconductors looks for a possible source of relatively stronger
attraction so that the formation of Cooper type pairs of charge
carriers becomes possible.  But we make no such argument and
discover the real interaction from our theoretical analysis.

\bigskip
In a recent conference [40], we presented our approach of
macro-orbital representation of a conduction electron to lay the
foundations of a viable theory of superconductivity.  We
discovered different aspects of this representation, for the first
time, in our recent study of the wave mechanics of two {\it hard
core} (HC) identical particles in 1-D box [41] and used them to
understand the unification of the physics of widely different many
body systems of interacting bosons and fermions [42] and reveal
the ground state of $N$ HC quantum particles in 1-D box [43].
While our approach also concludes Cooper type pairs as the origin
of superconductivity but it discovers some untouched aspects of
the electron-phonon interaction responsible for the formation of
such pairs.
        
\bigskip
The paper has been arranged as follows.  The Hamiltonian of the
electron fluid in a solid has been analyzed in Section 2.0 to
identify its universal component ($H_o(N)$, Eqn. 2, below),
-independent of the specific aspects of a superconductor or
a class of superconductors, while
the wave mechanics of a pair of conduction electrons, found to
serve as the basic unit of the fluid, has been examined in
Section 3.0 to conclude its several important aspects and to
discover that a conduction electron is better represented by a
macro-orbital ({\it a kind of pair waveform} as described in
Section 3.4.7) rather than a plane wave.  A wave function
that represents a general state of the electron fluid has
been constructed in Section 4.0 by using $N$ macro-orbitals for
$N$ conduction electrons and used to conclude their ground
state configuration.  While the equation of state of the
electron fluid has been analyzed to obtain the free energy
in Section 5.0, different aspects related to superconductivity,
such as criticality of electron fluid, onset of lattice strain,
energy gap and formation of ({\bf q}, -{\bf q}) bound pairs,
transition temperature, {\it etc.} are discussed in Section
6.0.  The paper is summed up by examining the consistency of
our model with other well known models such as BCS theory,
two fluid theory, $\Psi-$theory, {\it etc.} in Section 7.0
and making certain important remarks in Section 8.0.

\bigskip
For the first time, this paper identifies mechanical
strain in the lattice (produced by the zero-point force of
the conducting electrons) as {\it the main factor} responsible
for electron-phonon interaction leading to superconductivity;
this strain is different from the electrical strain ({\it i.e.}
electrical polarization of the lattice constituents produced
by electron charge) emphasized in BCS theory.  We believe that
electrical strain possibly adds to above mentioned
electron-phonon interaction.  Since the {\it zero-point force}
is a consequence, purely, of the wave nature of a quantum
particle like electron, the onset of the said mechanical strain
below certain temperature ({\it cf.} Section 6.2) rightly
represents a universal aspect of superconductivity.  It is
interesting to note that recent experimental studies confirm the
occurrence of lattice strain [44] and corroborate the fact that
phonons have major role in the mechanism of superconductivity
even in HTC systems [45].  We note that several theoretical
studies [46] relate charge fluctuation, spin fluctuation,
phase fluctuation, superconducting density fluctuation and/or
similar other factors with the onset of superconductivity, while
the present analysis sees a possibility of their coupling
with the mechanical strain in the lattice which serves as
the basic order parameter of the transition.

\bigskip
\centerline{\bf 2.0 Important Aspects of The Electron Fluid}

\bigskip
\noindent
2.1. {\bf Hamiltonian}

\bigskip
The hamiltonian of $N$ conduction electrons can be expressed,
to a good approximation, as
$$H(N) = -\frac{\hbar^2}{2m}\sum_i^N\bigtriangledown_i^2 +
\sum_{i<j}V(r_{ij}) + V'(N), \eqno(1) $$

\noindent
where $m$ is the mass of an electron, $V(r_{ij})$ is the
{\it central force potential} experienced by two electrons
and $V'(N)$ stands for the sum of all possible interactions
such as electron-phonon, spin-spin, spin-lattice, {\it etc.}
We assume that different components of $V'(N)$ can be treated
as perturbation on the states of
$$H_o(N) = H(N) - V'(N).  \eqno(2)$$  

\noindent 
which we identify as a universal component ({\it i.e.},
independent of the specific nature of a chosen superconductor or
a class of superconductors) of $H(N)$.  This breakup has an
advantage that the impact of {\it one} (or more than one)
component(s) of $V'(N)$ present in a chosen superconductor (or a
class of superconductors) can be examined as a perturbation on
the states of $H_o(N)$.  To find the states of $H_o(N)$, we
assume that the electron fluid is a Fermi fluid where
$V(r_{ij})$ is the sum of a short range strong repulsion
$V^R(r_{ij})$ and an {\it indirectly induced} weak attraction
$V^A(r_{ij})$ of slightly longer range.

\bigskip
To a good approximation, $V^R(r_{ij})$ can be equated to a
{\it hard core} (HC) interaction $V_{HC}(r_{ij})$ defined by
$V_{HC}(r_{ij} < \sigma) = \infty$ and  $V_{HC}(r_{ij} \ge
\sigma) = 0$ where $\sigma$ is the HC diameter of an electron.
One finds enough reasons to justify $V^R(r_{ij})\approx
V_{HC}(r_{ij})$, {\it e.g.}: (i) the screening effect of
the lattice greatly reduces the strength and range of
inter-electron repulsion, (ii) the conduction electrons
flow through certain types of channels {\it viz.}, a cylindrical
tube of diameter $d_c$ in the lattice or a 2-D slot of width $d_c$
between two parallel atomic planes; $d_c$ being a small fraction
of lattice constant $a$ speaks about the smallness of their
$\sigma$, and (iii) the density of conduction electrons
renders inter-electron distance ranging from a vaue
$> a/2$ ({\it cf.} Section 3.4.5) in systems with their
number density higher than that of atoms/molecules) to a
couple of $a$ in systems (where the said density is
lower) and this indicates about the smallness of the
range of $V^R(r_{ij})$.

\bigskip
Since no conduction electron comes out of a solid unless
a definite amount of energy ($\approx$ work function) is
supplied from outside, there are certain factors, such as
polarizability of lattice constituents, presence of $+ve$
charges in the background of mobile electrons, {\it etc.},
which bind each electron with the entire system (the lattice +
conduction electrons) indicating the presence of $V^A(r_{ij})$
({\it i.e.} the electron-lattice interaction leading to an
indirect inter-electron attraction). To a good approximation,
$V^A(r_{ij})$ can be replaced by a constant negative potential
$-V_o$ whose main role is to keep electrons within the volume
of the conductor.  This indicates that each conduction electron,
to a good approximation, can be identified as a freely moving HC
particle on the surface of a constant $-ve$ potential.

\bigskip
\noindent
2.2. {\bf  Basic unit of the fluid}

\bigskip
In what follows from the above discussion, the motion of each
conduction electron, to a good approximation, can be expressed
by a plane wave
$$u_{\bf p}({\bf b} ) = A\exp(i{\bf p}.{\bf b}) \eqno(3)$$ 

\noindent
where {\bf p} and {\bf b}, respectively, represent the momentum
(in wave number) and position vectors of an electron. However,
the plane wave motion is modified by its collision with other
electron(s) or the lattice structure.  A collision could either
be a two body collision (electron-electron collision), or a many
body collision ({\it e.g.}, two mutually colliding electrons also
collide simultaneously with other electron(s) or lattice
structure).  In the former case two electrons (say, e1 and e2)
simply exchange their momenta ${\bf p}_1$ and ${\bf p}_2$ or
positions ${\bf b}_1$ and ${\bf b}_2$ without any difference in
the sum of their pre- and post-collision energies. However, in the
latter case e1 and e2 could be seen to jump from their state of
${\bf p}_1$ and ${\bf p}_2$ to that of different momenta
${\bf p}'_1$ and ${\bf p}'_2$ (possibly of different energy) but
it is clear that to a good approximation they remain in one of
the possible states of two HC particles moving in the absence of
other electron(s) and/or lattice.  This implies that the complex
dynamics of the electron fluid can be described to a good
approximation in terms of the simple dynamics of a pair of HC
particles (discussed in Section-3 below) as its {\it basic unit}.

\bigskip
\centerline{\bf 3.0 Dynamics of Two HC Particles}  

\bigskip
\noindent
3.1. {\bf  Schr\"{o}dinger equation}

\bigskip
The Schr\"{o}dinger equation of two HC impenetrable particles
can be described by
$$\left(-\frac{\hbar^2}{2m}\sum_i^2\bigtriangledown^2_i +
V_{HC}(r)\right)\psi{(1,2)} = E(2)\psi{(1,2)} \eqno(4)$$
 
\noindent
While the dynamics of two electrons in a many body collision
involving lattice structure can be expected to encounter an
interaction different from that involved in a two electron or
many electron collision but the fact, that the end result of
any such collision is to take two electrons from their state
of ${\bf p}_1$ and ${\bf p}_2$ to that of ${\bf p}'_1$ and
${\bf p}'_2$, indicates that the difference is {\it unimportant}
and we can proceed with our analysis of $\psi{(1,2)}$ and use
its results to find how each electron in $\psi{(1,2)}$ state
assumes a phonon induced inter-electron attraction seemingly
essential for the occurrence of superconductivity in widely
different superconductors.

\bigskip
The process of solving Eqn. 4 is simplified by using: (i)
$V_{HC}(r) \equiv A\delta{(r)}$ where $A$ representing the
strength of Dirac delta function $\delta{(r)}$ is such that
$A \to \infty$ when $r \to 0$ (this type of equivalence has been
mathematically demonstrated by Huang [47] and physically
argued in Section 3.2, below), and (ii) the
{\it center of mass} (CM) coordinate system defined by,
$${\bf r}= {\bf b}_2 - {\bf b}_1 \quad {\rm and} \quad
{\bf k} = {\bf p}_2 - {\bf p}_1 = 2{\bf q}, \eqno(5)$$

\noindent
with {\bf r} and {\bf k}, respectively, representing the
relative position and relative momentum
of two electrons, and  
$${\bf R}= ({\bf b}_2 + {\bf b}_1)/2  \quad {\rm and} \quad
{\bf K} = {\bf p}_2 + {\bf p}_1, \eqno(6)$$

\noindent
with {\bf R} and {\bf K}, similarly, referring to the position
and momentum of their CM.  Without loss of generality, Eqns.
5 and 6 also render
$${\bf p}_1 = - {\bf q} + \frac{\bf K}{2} \quad {\rm and}
\quad {\bf p}_2 = {\bf q} + \frac{\bf K}{2}.
\eqno(7)$$
 
\noindent
By using these equations, one may express Eqn. 4 as     
$$\left(-\frac{\hbar^2}{4m}\bigtriangledown^2_R
-\frac{\hbar^2}{m}\bigtriangledown^2_r +
A\delta{(r)}\right)\Psi{(r,R)} = E(2)\Psi{(r,R)} \eqno(8) $$

\noindent
with 
$$\Psi{(r,R)} = \psi_k(r)\exp(i{\bf K}.{\bf R}).    \eqno(9) $$

\noindent
We note that the HC interaction affects only
the relative motion of two particles [$\psi_k(r)$] which
represents a solution of
$$\left(-\frac{\hbar^2}{m}\bigtriangledown^2_r +
A\delta{(r)}\right)\psi_k(r) = E_k\psi_k(r) \eqno(10) $$

\noindent
with $E_k = E(2)-\hbar^2K^2/4m$, while the CM motion
$[\exp(i{\bf K}.{\bf R})]$ remains unaffected. 

\bigskip
\noindent
3.2. {\bf Basis for} $V_{HC}(r) \equiv A\delta{(r)}$

\bigskip
The physical basis for $V_{HC}(r) \equiv A\delta{(r)}$ can be
understood by examining the possible configuration of two HC
particles (say P1 and P2) right at the instant of their
collision.  When P1 and P2 during a collision have their
individual CM located, respectively, at ${\rm r}_{CM}(1) =
\sigma/2$ and ${\rm r}_{CM}(2) = -\sigma/2$ (with ${\rm r}_{CM}$
being the distance of the CM of a particle from the CM of P1 and
P2), they register their physical touch at $r = 0$ and their
encounter with $V_{HC}(r)$ is a result of this touch beyond which
two HC particles can not be pushed in.  The process of collision
only identifies this touch; it does not register how far are the
CM points of individual particles at this instant. In other words
the rise and fall of the potential energy of P1 and P2 during
their collision at $r = 0$ is independent of their $\sigma$ and
this justifies $V_{HC}(r) \equiv A\delta{(r)}$.
It may, however, be mentioned that this equivalence will not be
valid in accounting for certain physical aspects of the system
({\it e.g.}, the volume occupied by a given number of particles)
where the real size of the particle assumes importance.

\bigskip
\noindent
{\bf 3.3. Statefunctions}

\bigskip
In order to find the statefunction $\Psi{(r,R)}$, -a solution of
Eqn. 8, we treat $A\delta{(r)}$ as a step potential.   Since
P1 and P2 experience zero interaction in the region $r \not = 0$,
they can be represented, independently, by plane waves except
around $r = 0$ where $A\delta{(r)} = \infty$.  However, in view
of the possible superposition of two waves, the state of P1 and
P2 can, in principle, be described by
$$\Psi{(1,2)}^{\pm} =
\frac{1}{\sqrt{2}}[u_{{\bf p}_1}({\bf r}_1)u_{{\bf p}_2}({\bf r}_2)
\pm u_{{\bf p}_2}({\bf r}_1)u_{{\bf p}_1}({\bf r}_2]. \eqno(11)$$

\noindent
But we find that $\Psi{(1,2)}^+ $ (of $+ve$ symmetry for the
exchange of two particles) does not represent the {\it desired}
wave function of two HC particles since, as required, it does
not vanish at ${\bf r}_1 = {\bf r}_2$ where $A\delta{(r = 0)} =
\infty$, while the other function $\Psi{(1,2)}^-$ of $-ve$
symmetry has no such problem.  We addressed this problem in our
recent analysis of the 1-D analogue of Eqn. 8 in relation to our
detailed study of the wave mechanics of two HC impenetrable
particles in 1-D box [41].  In what follows from this study [41]
one may easily find that the state of two such particles can be
expressed by
$$\zeta{(r,R)}^{\pm} =
\zeta_k(r)^{\pm}\exp{(i{\bf K}.{\bf R})}   \eqno(12)$$

\noindent
with 
$$\zeta_k(r)^- = \sqrt{2}\sin{({\bf k}.{\bf r}/2)} \eqno(13)$$ 

\noindent
of $-ve$ symmetry, and 
$$\zeta_k(r)^+ = \sqrt{2}\sin{(|{\bf k}.{\bf r}|/2)} \eqno(14)$$

\noindent
of $+ve$ symmetry for the exchange of their ${\bf r}_1$ and
${\bf r}_2$ (or ${\bf k}_1$ and ${\bf k}_2$).  It is obvious
that in a given state of the pair only ${\bf r}$ is variable;
any change in ${\bf k}_1$ and ${\bf k}_2$ and/or the angle
between ${\bf k}$ and ${\bf r}$ implies a change in the state.
We note that the second derivative of $\zeta_k(r)^+$ with
respect to $r$ has $\delta-$like singularity at $r = 0$ which,
however, can be reconciled for the presence of infinitely
strong repulsive potential at $r = 0$.

\bigskip
\noindent
{\bf 3.4 Characteristic Aspects}

\bigskip
\noindent
3.4.1. {\it Nature of relative motion :}  We note that
$\zeta_k(r)^{\pm}$, describing the relative motion of two
HC particles, is a kind of {\it stationary matter wave}
(SMW) which modulates the probability $|\zeta_k(r)^{\pm}|^2$
of finding two particles at their relative phase position
$\phi = {\bf k}.{\bf r}$ in the $\phi-$space.
Interestingly, the equality $|\zeta_k(r)^-|^2 =
|\zeta_k(r)^+|^2$ concludes an {\it important fact} that the
relative configuration and relative dynamics of two HC particles
is independent of their fermionic or bosonic nature.  This
implies that the requirement of fermionic symmetry of
electrons should be enforced on the wave functions of their
${\bf K}-$motions or spin motions and we use this inference in
constructing $N-$electron wave function in Section-4.  In
agreement with Eqn. 7, the SMW character of $\zeta_k{(r)}^{\pm}$
also reveals that two HC particles in $\zeta{(r,R)}^{\pm}$ state
have equal and opposite momenta ({\bf q},-{\bf q}) in the frame
attached to their CM which moves with momentum ${\bf K}$ in the
laboratory frame. It is also evident that the two particles in
their relative motion maintains a center of symmetry at their CM
(the point of their collision) which implies
$${\bf r}_{CM}(1) = -{\bf r}_{CM}(2) =
\frac{\bf r}{2} \quad {\rm and} \quad {\bf k}_{CM}(1)
= -{\bf k}_{CM}(2) = {\bf q} \eqno(15)$$

\noindent
where ${\bf r}_{CM}$(i) and ${\bf k}_{CM}$(i), respectively,
refer to the position and momentum of $i-$th particle with
respect to the CM of two particles.
 
\bigskip
\noindent
3.4.2. {\it MS and SS states :}  Since $\zeta{(r,R)}^{\pm}$ is
a result of the superposition of two plane waves of momenta
${\bf p}_1$ and ${\bf p}_2$ and the state is basically an
eigenstate of the momentum/energy operators of the relative
and CM motions of two particles in superposition ({\it not of
individual particles}), it could be rightly identified as a
state of {\it mutual superposition} (MS) of two particles.
However, one may have an alternative picture by presuming
that each of the two
particles after their collision falls back on the pre-collision
side of $r = 0$ (the point of collision) and assumes a kind
of {\it self superposition} (SS) ({\it i.e,}, the superposition
of pre- and post-collision states of one and the same particle).
Interestingly, this is also described by $\zeta{(r,R)}^{\pm}$
because it also represents a superposition of the plane wave
of momentum ${\bf p}_1$ (the pre-collision state of P1) and a
similar wave of momentum ${\bf p}'_1 =  {\bf p}_2$ representing
post-collision state of P1 because two particles exchange
their momenta on their collision; the same effect can be
seen with P2.  However, since P1 and P2 are identical
particles and we have no means to ascertain whether they
exchange their positions or bounce back after exchanging
their momenta, we can use $\zeta{(r,R)}^{\pm}$ to identically
describe the MS state of two particles or the SS states of
individual particle and this helps us in developing the
{\it macro-orbital representation} of each electron in the
fluid ({\it cf.}, point 3.4.7).

\bigskip
\noindent
3.4.3. {\it Values of $<r>$, $<\phi>$ and $<H(2)>$ :}  The SMW
waveform, $\zeta_k(r)^{\pm}$, has series of antinodal regions
between different nodal points at $r = \pm n\lambda/(2\cos\theta)$
(with $n$= 0,1,2,3, ... and $\theta$ being the angle between
{\bf q} and {\bf r}).  This implies that two particles can
be trapped on the $r$ line {\it without disturbing their energy or
momenta} by suitably designed cavity of impenetrable infinite
potential walls.  For example, one may possibly use two pairs of
such walls and place them at suitable points perpendicular to
${\bf k}_1$ and ${\bf k}_2$ or to the corresponding ${\bf k}$
and ${\bf K}$).  In case of ${\bf k} || {\bf r}$ (representing a
$s-$wave state) one can use a cavity of only two such walls placed
at the two nodal points located at equal distance on the opposite
sides of the point ($r=0$) of their collision.  Using the fact
that the shortest size of this cavity can be only $\lambda$, we
easily find
$$<r>_o \quad = \frac{<\zeta_k(r)^{\pm}|r|\zeta_k(r)^{\pm}>}
{<\zeta_k(r)^{\pm}||\zeta_k(r)^{\pm}>} =
\frac{\lambda}{2}  \eqno(16)$$

\noindent
as the shortest possible $<r>$.  To this effect, integrals are
performed between $r=0$ (when the two particles are at the center
of cavity) to $r = \lambda$ (when one particle reaches at
$r = \lambda/2$ and the other at $r = -\lambda/2$ representing
the locations of the two walls which reflect the particles back
inside the cavity).  Following a similar analysis for the general
case we identically find $<r> \quad = \lambda/(2\cos\theta)$
which not only agrees with Eqn. 16 but also reveals that the two
particles assume $<r> = <r>_o$ only when they have head-on
collision.  Evidently, from an experimental view point, two HC
particles never reach closer than $\lambda/2 = \pi/q$ and in this
situation their individual locations ({\it cf.} Eqn. 15) are
given by $<{\bf r}_{CM}(1)>_o = - <{\bf r}_{CM}(2)>_o =
\lambda/4$. Finding similar result for their shortest possible
distance in $\phi-$space and $<V_{HC}(r)>$, {\it etc.} we note
that $\zeta_k(r)^{\pm}$ state is characterized by
$$<\zeta_k(r)^{\pm}|r|\zeta_k(r)^{\pm}> \quad \ge \quad
\lambda/2 \quad {\rm and} \quad <\psi_k(r)^{\pm}|\phi
|\psi_k(r)^{\pm}> \quad  \ge 2\pi,  \eqno(17)$$
$$<\zeta_k(r)^{\pm}|V_{HC}(r)|\zeta_k(r)^{\pm}>
\quad = \quad <\zeta_k(r)^{\pm}|A\delta{(r)}|\zeta_k(r)^{\pm}>
\quad = 0, \eqno(18)$$
$$E(2)   = \quad <\zeta{(r,R)}^{\pm}|H(2)|\zeta{(r,R)}^{\pm}>
\quad = \left[\frac{\hbar^2k^2}{4m} +
\frac{\hbar^2K^2}{4m}\right]. \eqno(19) $$
  
\noindent
It is evident from these equations that two particles in
$\zeta{(r,R)}^{\pm}$, basically, have kinetic energy.
We note that the fact, that SMW state (({\it cf.} Section 3.4.1)
does not allow two HC particles to occupy a common point in
real space (two particles always stay on the opposite sides
of their CM), provides a strong basis to justify
Eqn. 18.  We analyze Eqn. 18 for its general validity in
Appendix-A and we find that it is valid for all physically
relevant situations of two particles.  We also note that
$\zeta{(r,R)}^{\pm}$ is not an eigenstate of the
momentum/energy operators of individual particle.  In stead,
it is the eigenstate of the energy operator of the
pair of particles which, naturally, share $E(2)$ equally.

\bigskip
\noindent
3.4.4. {\it quantum size :} In what follows from Eqns. 17 and 18,
a HC particle of momentum $q$ exclusively occupies $\lambda/2$
space if $\lambda/2 > \sigma$ because only then the two particles
maintain $<r> \ge \lambda/2$.  We call $\lambda/2$ as quantum
size of the particle.  One may also
identify quantum size as the size of a particle (say P1) as
seen by the other particle (say P2) or {\it vice versa} in
$\zeta_k(r)^{\pm}$ state of their wave superposition.  To this
effect we may consider P1 as an object to be probed and P2 as a
probe (or {\it vice versa}) and apply the well known principle
of image resolution.  We find that P2 can not resolve the $\sigma$
size of P1 if $\lambda/2 > \sigma$ and the effective size of P1 as
seen by P2 (or {\it vice versa}) would be limited to $\lambda/2$.
But the situation is different for the particles of $\lambda/2
\le \sigma$ because here they can resolve the $\sigma$ size of
each other and P1 and P2 would see each other as particles of
size $\sigma$ in all states of $q \ge h/2\sigma$.  This concludes
that the effective size of low momentum particles ($q <
h/2\sigma$) is $q-$dependent, while that of high momentum
particles ($q \ge h/2\sigma$) is $q-$independent;  this renders
an important aspect which can explain why many body systems
exhibit the impact of wave nature only at low temperatures.

\bigskip
On the qualitative scale our meaning of $``$quantum size$"$ seems
to be closer to what Huang [47] refers as $``$quantum spread$"$
but on the quantitative scale, while we relate $``$quantum
size$"$ of a particle with its momentum by a definite relation
$\lambda/2 = \pi/q$, $``$quantum spread$"$ has not been so
related.  The fact, that no particle can be accommodated in
a space shorter than $\lambda/2$, implies that $``$quantum
size$"$ could be identified as the minimum of quantum spread of
a particle or as the minimum possible size of space exclusively
occupied by the particle. It may also be mentioned that our
meaning of $``$quantum size$"$ differs from that of {\it
quantum size} word in $``$quantum size effects$"$ on the
properties of thin films, and small clusters of atoms [48],
{\it etc.}

\bigskip
\noindent
3.4.5. {\it zero-point force :}
In what follows from the above discussion, each HC particle
in a fluid exclusively occupies a minimum space of size
$\lambda/2$ which, obviously, increases with fall in $T$.
Evidently, at certain $T = T_o$, at which the average
$\lambda/2$ equals $d$ ({\it the average nearest neighbor
distance}), almost
all particles have their minimum possible spread and they find
themselve in the ground state of a box of size $d$ (a cavity
formed by neighboring atoms) with their momenta frozen at
$q = q_o = \pi/d$.  Using the thermal de Broglie
wavelength $\lambda_T = h/\sqrt{2\pi mk_BT}$ as average
$\lambda$, we have
$$T_o = \frac{h^2}{8\pi mk_Bd^2}.  \eqno(20)$$     

\noindent
Evidently, each particle at $T \le T_o$ tends to
have $\lambda/2 > d$ by expanding the cavity size; for
this purpose it exerts its zero-point force
$f_o = h^2/4md^3$ against inter-particle force
($f_a$) that decides the size and structure of the
cavity.  This happens also to conduction electrons
constrained to move through narrow channels
({\it e.g.}, a cylindrical tube or 2-D slot
between two parallel lattice plains in the systems like HTC
superconductors) of diameter or width $d_c$ and they all
exert $f_o$ on the lattice against its $f_a$ deciding
$d_c$.  Consequently, lattice has non-zero mechanical strain
which plays a crucial role for superconductivity
({\it cf.} Section 6.0).  In this cotext an estimate of
shortest $d = (v/n)^{1/3}$ (with $v$ = unitcell volume
per atom and $n$ = number of conduction electrons cotributed
by the atom) renders $d = a/2$ by using $n = 8$ (the maximum
possible $n$).  Since even this $d$ is larger than an
expected $d_c$, the lowest possible $q = q_o$ for an
electron should be decided by $d_c$ which means that
$d_c$ is more relevant than $d$ to determine the ground
state properties of electron fluid.

\bigskip
\noindent 
3.4.6. {\it Phase correlation : } Applying the standard
procedure [49, 50] for determining the quantum correlation
potential $U(\phi)$ between two particles of a many body
system, we easily find that two particles have a
$\phi-$correlation defined by
$$U(\phi) = -k_BT\ln{|\zeta_k(r)^{\pm}|^2}
=  -k_BT\ln{[2\sin^2(\phi/2)]}  \eqno(21)$$
\noindent
which represents a kind of binding between them in the
$\phi-$space.  We note that $U(\phi)$ has a series of
$\phi-$points at which it has minimum value (= $-k_BT_o\ln{2}$
at $\phi = (2n+1)\pi$ with $n$ being an integer) and maximum
value (= $\infty$ at $\phi = 2n\pi$) which implies that
wavemechanical superposition of two particles tries to arrange
them in $\phi-$space at the points of minimum $U(\phi)$
which are separated by $\Delta\phi = 2n\pi$.
The experimentally observed coherence in the motion of
electrons particularly in their superconducting state is a
consequence of this fact.  It may be mentioned that $T$ in
Eqn. 21 should be replaced by $T_o$ (Eqn 20 with $d = d_c$)
representing $T$ equivalent of $\varepsilon_o = h^2/8md_c^2$
because $q-$motion energy of each electron at $T \le T_o$
gets frozen at $\varepsilon_o$ ({\it cf.} Section 4.2).

\bigskip
\noindent
3.4.7. {\it Macro-orbital representation :}  We note that
in spite of their binding in the $\phi-$space as concluded
above, two HC particles in the real space experience a kind of
mutual repulsion, if they have $<r> < \lambda/2$, or no force,
if $<r> \ge \lambda/2$.  This implies that each particle in
$\zeta{(r,R)}^{\pm}$ pair state can be identified as
independent particle in its self superposition ({\it cf.}
point 3.4.2) represented by a kind of pair waveform
$\xi \equiv \zeta^{\pm}(r,R)$ proposed to be known as
macro-orbital and expressed as,
$$\xi_i = \sqrt{2}\sin[({\bf q}_i.{\bf r}_i)]
\exp({\bf K}_i.{\bf R}_i), \eqno(22)$$
\noindent
where $i$ ($i$ = 1 or 2) refers to one of the two particles; here
$r_i$ could be identified with $r_{CM}(i)$ ({\it cf.} Eqn. 15)
which varies from $r_i = 0$ to $r_i = \lambda/2$, while $R_i$
refers to the CM point of $i-$th particle.  Although, two
particles in $\zeta{(r,R)}^{\pm}$ state are independent but it
is clear that each of them represents a ({\bf q}, -{\bf q}) pair
whose CM moves with momentum {\bf K} in the lab frame.  This
implies that each particle in its macro-orbital representation
has two motions: (i) the plane wave $K-$motion which remains
unaffected by inter-particle interactions, and (ii) the
$q-$motion, affected by the inter-particle interaction.  In other
words a macro-orbital identifies each electron as a particle
of effective size $\lambda/2$ moving with momentum $K$ and this
gives due importance to the quantum size of a quantum particle or
equivalently to the wave packet (again of size $\lambda/2$)
manifestation of a quantum particle as invoked by wave mechanics.
We find that this picture is consistent with two fluid
phenomenology of superconductivity ({\it cf.} Section 7.2).
Since $\zeta{(r,R)}^{\pm}$ is neither an eigenfunction of the
energy operator nor of the momentum operator of a {\it single
particle}, each particle shares the pair energy $E(2)$ equally.
We have
$$E_1 = E_2 = \frac{E(2)}{2} = \frac{\hbar^2q^2}{2m} +
\frac{\hbar^2K^2}{8m}  \eqno(23) $$

\noindent
It is interesting to note that two particles, having different
momenta (${\bf p}_1$ and ${\bf p}_2$) and different energy
before their superposition, have equal momentum/energy in
$\zeta{(r,R)}^{\pm}$ state which indicates that their wave
superposition take them into a kind of degenerate state and this
tends to happen with all electrons when the system is cooled
through certain $T = T_a$ lower than $T_o$ ({\it cf.} Section
6.1).  In order to show that $\xi_i$ fits as a solution of
Eqn. 4 [with $V_{HC}(r) \equiv A\delta (r)$], we recast the
two particle hamiltonian
$H_o(2) = -\sum_i^2(\hbar^2/2m)\bigtriangledown^2_i +
A\delta{(r)}$ as $H_o'(2) = \sum_i^2h(i) + A\delta{(r)}$ by
defining
$$h_i = -\frac{\hbar^2}{2m}\bigtriangledown^2_i \quad
{\rm and} \quad h(i) = \frac{h_i + h_{i+1}}{2} =
-\frac{\hbar^2}{8m}\bigtriangledown^2_{R_i}
-\frac{\hbar^2}{2m}\bigtriangledown^2_{r_i} \eqno(24)$$

\noindent
with $h_{N+1} = h_1$ for a system of $N$ particles.  It is
evident that $\xi_i$ is an eigenfunction of $h(i)$ with $<h(i)>
= (\hbar^2q_i^2/2m + \hbar^2K_i^2/8m)$ and the two particle
wave function, $\Phi (2) = \xi_1\xi_2$ (or with added permuted
terms), is an eigenfunction of $H_o'(2)$ with $<H_o'(2)> =
E(2)$ ({\it cf.} Eqn. 19) because $<A\delta{(r)}> =
A|\xi_1|^2_{r_1=0}|\xi_2|^2_{r_2=0} = A\sin^2q_1r_1|_{r_1=0}
\sin^2q_2r_2|_{r_2=0} = 0$ since $r=0$ implies $r_1 = r_2 = 0$
({\it cf.} Eqn. 15).

\bigskip
\noindent
3.4.8. {\it Accuracy and relevance of macro-orbitals :}  While
the fact, that the fall of an electron into its SS state
({\it cf.}, Section 3.4.2) is independent of the details of the
collision ({\it i.e.}, two body collision, many body collision
or the collision with the lattice structure), justifies its
representation by $\xi_i$ in general, we also find that the
functional nature of $\xi_i$ matches almost exactly with
$$\eta_{q,K}(s,Z) =
A\sin[({\bf q}.{\bf s})]\exp({\bf K}.{\bf Z}) \eqno(25)$$

\noindent
representing a state of a particle in a cylindrical channel
with {\bf s} being the 2-D space vector perpendicular to
$z-$axis (the axis of the channel) and,
$$\eta_{q,K}(z,S) =
B\sin[({\bf q}.{\bf z})]\exp({\bf K}.{\bf S})  \eqno(26)$$

\noindent
which represents a similar state of a particle trapped between
two parallel impenetrable potential sheets. Interestingly,
since superconductivity is a behavior of low energy electrons
and a conduction electron in a solid can be visualized,
to a good approximation, as a particle moving along the axis of
cylindrical channel ({\it e.g.} in a conventional superconductor)
or that moving between two parallel atomic sheets ({\it e.g.} in
HTC systems), the accuracy and relevance of macro-orbitals in
representing the conduction electrons in their low energy states
is well evident.

\bigskip
\centerline{\bf 4.0  States of $N-$Electron Fluid}

\bigskip
\noindent
{\bf 4.1 General state}

\bigskip
Using $N$ macro-orbitals for $N$ electrons and following standard
method, we have
$$\Psi^j_n(N) = \Pi_i^N\zeta_{q_i}(r_i)
\sum_P^{N!} (\pm 1)^P\Pi_i^N\exp{[i(P{\bf K}_i{\bf R}_i)}]
\eqno(27)$$

\noindent
for one of the $N!$ microstates of the system of energy $E_n$
({\it cf.} Eqn. 29, below).  Here $\sum_P^{N!}$ represents the
sum of $N!$ product terms obtainable by permuting $N$ particles
on different ${\bf K}_i$ states with $(+1)^P$ and $(-1)^P$,
respectively, used for selecting a symmetric and anti-symmetric
wave function for an exchange of two particles.  In principle,
the permutation of $N$ particles on different ${\bf q}_i$ states
renders $N!$ different $\Psi^j_n(N)$ and we have
$$\Phi_n(N) = \frac{1}{\sqrt{N!}}\sum_j^{N!}\Psi^j_n(N)
\eqno(28)$$

\noindent
as the complete wave function of a possible quantum state of
energy $E_n$ given by
$$E_n = \sum_i^N\left[\frac{\hbar^2q_i^2}{2m}  +
\frac{\hbar^2K_i^2}{8m}\right]  \eqno(29)$$

\noindent
where $q_i$ and $K_i$ can be an integer multiple of $\pi/d_c$
and $\pi/L$, respectively.  To follow Eqn. 29, one may use
Eqn. 24 to recast $H_o(N) \approx \sum_i^Nh_i +
\sum_{i>j}^NA\delta{(r_{ij})}$ as
$$H_o(N) =
\sum_i^Nh(i) + \sum_{i>j}^NA\delta{(r_{ij})} \eqno(30)$$

In what Follows from Eqn. 29, the energy of conduction
electrons is basically kinetic which seems to imply that
these electrons constitute a system of some kind of
non-interacting fermions, while this is not true.  This
{\it apparent result} is obtained because
$<V_{HC}(r_{ij})> = 0$ ({\it cf.} Eqn. 18) which agrees with
the fact that two electrons do not occupy common point
in real space.  In addition one may find that the presence of
$V_{HC}(r_{ij})$ restricts $<r>$ through $<r> \ge \lambda/2$
(Eqn. 17) (or $q$ through $q \ge q_o$) which clearly indicates
that $V_{HC}(r_{ij})$ plays an important role in deciding the
relative configuration ({\it i.e.} the allowed $<r>$, $<\phi>$
and $q$) of conduction electrons, particularly, when the system
assumes the ground state of their $q-$motions with all
electrons having $q = q_o$.

\bigskip
\noindent
{\bf 4.2 Ground state}

\bigskip
We note that each conduction electron has two motions $q$ and
$K$.  While the $q-$motions are constrained to have $q \ge q_o
(=\pi/d_c)$, -the lowest possible $q$ of a particle restricted
to move through channels of size $d_c$, the $K-$motions are
guided by the Pauli exclusion principle.  Consequently, the
ground state of the fluid is defined by all $q_i = q_o$ and
different $K_i$ ranging between $K=0$ to $K=K_F$ (the Fermi wave
vector) which render
$$E_{GSE} = N\varepsilon_o + \bar{E_K} = N\frac{h^2}{8md_c^2} +
\frac{1}{4}.\frac{3}{5}NE_F   \eqno(31)$$

\noindent
as the {\it ground state energy} of the fluid.  Here
$\varepsilon_o = h^2/8md_c^2$ represents lowest possible energy
of the $q-$motion of an electron and $\bar{E_K}$ being the net
$K-$motion energy of $N$ electrons with
$E_F$ being the Fermi energy; the factor 1/4 in the last term
represents the fact that each electron in its macro-orbital
representation behaves like a particle of mass $4m$ for its
$K-$motions.  In order to understand how different inter-particle
interactions enter in our formulation to control the ground state
energy of electron fluid, it is important to note that $d_c$ in a
given system is decided by all such interactions.  Naturally, all
these interactions indirectly control the ground state momentum
through $q_o = \pi/d_c$ and hence the ground state energy
$\varepsilon_o$.  Expressing $E_{GSE}$ (Eqn. 31) in terms of its
temperature equivalent, we have
$$T_{GSE} = T_o + \bar{T(E_K)} \approx T_o + 0.15T_o  \eqno(32)$$
 
\noindent
where we use $T_o \equiv \varepsilon_o$ and $\bar{T(E_K)} \equiv
3E_F/20)$.  In writing $\bar{T(E_K)} = 0.15T_o$ we approximated
$E_F (\approx h^2/8md^2)$ to $\varepsilon_o (= h^2/8md_c^2)$ by
using $d_c$ for $d = ({\rm V}/N)^{1/3}$
where ${\rm V}$ is the net volume of
the solid containing $N$ electrons.  Since $d$ is always expected
to be larger than $d_c$, $T = 1.15T_o$ (Eqn. 32) can be identified
as the upper bound of $T_{GSE}$, while $T_o$ being the lower bound.

\bigskip
We note that: (i) each conduction electron in the ground
state is expressed by a stabilized macro-orbital, $\sin{(q_or)}
\exp{i({\bf K}.{\bf R})}$, for which we easily have $< k> =
<-i\hbar\bigtriangledown_r> = 0$ by using the fact that $r$
varies between $r = 0$ to $r = d_c$ and (ii) $<r>$ of each
conduction electron lies on the axis of the cylindrical tube
(the channel through which they move).  While inference-(i) 
concludes that for all practical purposes two conduction
electrons cease to have relative momentum indicating loss of
collisional motion or scattering with other electrons or
lattice (with $q = q_o$ simply representing the momentum of their
localization), inferences-(i and ii) reveal that conduction
electrons can move ({\it if they are set to move}) only in the
order of their locations in the channel(s), obviously,
with identically equal momentum, -a characteristic of coherent
motion.  Since the conduction
electrons are the constituents of the solid, their real space
positions (at least in the ground state) have to be compatible
with the crystal structure.

\bigskip
\centerline{\bf 5.0 Equation of State}

\bigskip
What follows from Eqn. 29, the energy of a particle in our
system can be express as
$$\epsilon = \varepsilon{(K)} + \varepsilon{(k)} = 
\frac{{\hbar}^2K^2}{8m} +
\frac{{\hbar}^2k^2}{8m}. \eqno(33)$$

\noindent
However, since the lowest $k = 2q$ is restricted to 
$2q_o$ for the condition, $q \ge q_o$, $\epsilon$ can have
any value between ${\varepsilon}_o = {\hbar}^2q_o^2/8m$
and $\infty$.  Interstingly, this possibility exists even
if ${{\hbar}^2k^2}/{8m}$ in Eqn. 33 is replaced by the
lowest energy ${\varepsilon}_o$ since $K$ can have any
value between 0 and $\infty$.  In other words, we can use
$$\epsilon = \frac{{\hbar}^2K^2}{8m} + {\varepsilon}_o
\eqno(34)$$

\noindent
which is valid, to a very good approximation, at low
temperatures where we intend to study the system. Using
Eqn. 34 in the starting expressions of the standard
theory of a system of fermions [51, Ch.8] we obtain
$$\frac{PV}{k_BT} = -{\Sigma}_{\varepsilon{(K)}}
\ln{[1 + z\exp{(- \beta[{\varepsilon{(K)} +
{\varepsilon}_o}]})}] \eqno (35)$$

\noindent
and
$$N = {\Sigma}_{\varepsilon{(K)}}\frac{1}{z^{-1}\exp{(\beta
[{\varepsilon{(K)} + {\varepsilon}_o}])} + 1}  \eqno (36)$$

\noindent
with $\beta = \frac{1}{k_BT}$ and fugacity 
$$z=\exp{({\beta}{\mu})} \quad \quad (\mu = {\rm chemical
\quad potential}). \eqno(37)$$

\noindent
Once again, by following the steps of the standard theory
[51] and redefining the fugacity by
$$z' = z\exp{(-\beta{\varepsilon}_o)} =
\exp{[\beta(\mu - {\varepsilon}_o)]} =
\exp{[\beta\mu']} \quad \quad {\rm with} \quad
\mu' = \mu - {\varepsilon}_o  \eqno(38)$$

\noindent
we easily have  
$$\frac{P}{k_BT} = - {\frac{2\pi{(8mk_BT)^{3/2}}}{h^3}}
{\int}_0^{\infty}x^{1/2}\ln{(1-z'e^{-x})}dx =
\frac{g}{{\lambda}^3}f_{5/2}(z') \eqno(39)$$

\noindent
and 
$$\frac{N}{V} = {\frac{2\pi{(8mk_BT)^{3/2}}}{h^3}}
{\int}_0^{\infty}\frac{x^{1/2}dx}{z'^{-1}e^x -1} =
\frac{g}{{\lambda}^3}f_{3/2}(z') \eqno(40)$$

\noindent
where $g$ is the weight factor that arises from inherent
character such as spin of particles,
$x = \beta{\varepsilon}(K)$, $\lambda_T =
h/(2{\pi}(4m)k_BT)^{1/2}$ and $f_n(z')$ has its usual
expression.  This reduces our problem of HC particles
to that of non-interacting fermions but with a difference.
We have $m$ replaced by $4m$ and $z$ by $z'$
(or $\mu$ by $\mu' = \mu - \varepsilon_o$).  The range of
$z$ and $z'$ remain unchanged.  In other words if $\mu$
and $z$ are, respectively, replaced by $\mu'$ and $z'$,
system of HC fermions can be treated as a system of
non-interacting fermions.  As such we can use
Eqns. 35 and 36 and Eqns. 39 and 40 to evaluate
different thermodynamic properties of our system.
For example, the internal energy
$U = - \frac{\partial}{\partial\beta}(\frac{PV}{k_BT})|_{z,V}$
of our system can be expressed as, 
$$U = \frac{3}{2}k_BT\frac{V}{{\lambda}^3}
f_{5/2}(z') + N\varepsilon_o = U' +
N\varepsilon_o  \eqno(41)$$

\noindent
with $U'= - \frac{\partial}{\partial\beta}
(\frac{PV}{k_BT})|_{z',V}$ being the internal energy
contribution of non-interacting quasi-particle fermions
representing $K-$motions and $N\varepsilon_o$
being the added contribution from $k-$motions.  Similarly, we
have Helmholtz free energy
$$A = N\mu - PV = N\varepsilon_o + (N\mu' - PV) =
N\varepsilon_o + A' \eqno(42)$$

\noindent
with $A'$ being the Helmholtz free energy of non-interacting
fermions.  Following standard methodology, we now analyze
the free energy $A$ for the physical conditions for which
it becomes critical and leads to superconductivity.

\bigskip
\centerline{\bf 6.0  Important Aspects of Superconductivity}

\bigskip
\noindent
{\bf 6.1  Free energy and its criticality}

\bigskip
The free energy of the electron fluid (Eqn. 42) is a sum of two
terms: (i) $A'$ representing the contribution of plane wave
$K-$motions which define a system of non-interacting
quantum quasi-particles of fermionic symmetry and mass $4m$
and (ii) $N\varepsilon_o$ representing the zero-point energy
of $q-$motions.  We note that a term like $A'$ alone can
represent our system at higher temperatures at which $\lambda$
of electrons at large satisfies $\lambda/2 << d_c$ for which
they do not have an effective wave superposition envisaged
for their macro-orbital representation and they can be
represented, to a good approximation, by plane waves.  This
observation agrees with the experimental fact that the behavior
of electron fluid in solids at high temperatures fits very well
with the theory of Fermi gas (non-interacting fermions)
available in every text on Fermi statistics, {\it e.g.}, [47, 51].
Since $A'$ does not become critical at any $T$, the criticality
of the electron fluid leading to its superconductivity must
arise only with $N\varepsilon_o = Nh^2/8md^2$.  However, since
$N\varepsilon_o $ has no explicit dependence on the parameters
such as temperature $T$, pressure $P$, {\it etc.}, it provides
no explicit mathematical solution for $T_c$, $P_c$, {\it etc.} at
which $N\varepsilon_o$ becomes critical.  It is for this reason
that {\it we examine the system for its criticality by analyzing
its evolution on cooling} and using the condition $q \ge q_o$
which makes the system critical when $q$ tend to fall below
$q_o$ at certain $T = T_a$.

\bigskip
Since $\lambda_T$ increases on cooling, almost all electrons
can have $\lambda/2 = d_c$ (or $q = q_o$) at $T_a$ close to
$T_o$ (Eqn. 20 with $d = d_c$).  However, Pauli exclusion
can be expected to push it down because $q = q_o$ state for
all electrons represents a kind of degenerate state which can
be disturbed effectively when electrons have enough energy in
their $K-$ motions ({\it cf.} Section 6.4).  Evidently,
to have a good estimate of $T_a$ we find the number of
conduction electrons $N_{2q_o}(T)$ in $q= 2q_o$ state
(the first excited state of $q-$motion) at $T = T_a$ by using
$N_{2q_o}(T) = N\exp{[-(4\varepsilon_o - \varepsilon_o)/k_BT]}$
as a good approximation.  We find $N_{2q_o}(T_o) \approx
10^{-1}N$ which means that only about $90\%$ particles
occupy $q = q_o$ state at $T = T_o$ indicating that
$T_a < T_o$.  However, a similar estimate of $N_{2q_o}(T)$,
for $T = 0.15T_o$ which represents the $T$ equivalent of
the least amount of $K-$motion energy retained by electrons
at $T = 0$, reveals $N_{2q_o}(0.015T_o) \approx 10^{-7}N$
which implies that  the percentage of electrons in $q = q_o$
state is $99.99999\%$.  Evidently, the occupancy of
$q = q_o$ state by macroscopically large number of electrons
seems to reach its completion {\it for all
practical purposes} at a $T > 0.15T_o$.   Naturally, electrons 
can have $q < q_o$ by using their zero-point force to
expand the channel size at $T = T_a$ which is expected to
fall between $T_o$ and $0.15T_o$.  We analyze this process
of straining the lattice in the follwing section.

\bigskip
\noindent
{\bf 6.2 Onset of lattice strain}

\bigskip
In what follows from the above analysis, electrons tend to
have $q < q_o$ ({\it i.e.} $\lambda/2 > d_c$) when they
are cooled through $T_a$.  They occupy more space
in the lattice structure by using their zero point force
$f_o = h^2/4md_c^3$ against the interatomic forces (say $f_a$)
which restore the lattice structure.  The equilibrium between
$f_o$ and $f_a$, obviously, renders a non-zero strain
$\Delta d = d_c' - d_c$ (with new $q = q_o' = \pi/d_c'$) in
the lattice and this happens for almost all conduction
electrons leading to an onset of the process of lattice
straining around $T_a$.  The experimental
fact, that liquids $^4He$ and $^3He$ on their cooling,
respectively, through $2.17$K and $0.6$K (matching closely with
$T_a$ [$0.15T_o < T_a < T_o$ for $T_o \approx 1.4$K) exhibit
volume expansion ({\it i.e.} -ve volume expansion coefficient)
[52], proves that $f_o$ (expected to operate around $T_a \le
T_o$) undoubtedly produces strain (expansion) in $He-He$ bonds
and there is no reason for which similar effect is not expected
from the $f_o$ of conduction electrons in superconductors.  In
fact the recent experimental studies [44, 45] have confirmed
the presence of mechanical strain in HTC systems.  

\bigskip
\noindent
{\bf 6.3 Energy gap and (q, -q) bound pairs}

\bigskip
With the onset of lattice strain $\Delta d$ ({\it cf.}
Section 6.2), the $q-$motion energy of a conduction electron
falls below $\varepsilon_o $ by 
$$\Delta\epsilon = \varepsilon_o  - \varepsilon_o' =
\frac{h^2}{8md_c^2} - \frac{h^2}{8m(d_c+\Delta d)^2} =
\frac{h^2}{4md_c^3}(\Delta d).  \eqno(43)$$

\noindent
In view of the nature of $q-$ and $K-$motions [{\it cf.}
Eqns. 8-10], it is evident that the process of straining the
lattice only affects the $q-$motions and $\Delta\epsilon$
represents a decrease in the zero-point energy of this motion
only.  A simple analysis of the equilibrium between $f_a$
({\it cf.} Section 6.2) and $f_o$, concludes [53] that, to a
good approximation, half of $\Delta\epsilon$ is {\it stored
with the lattice as its strain energy} ($\epsilon_s$) leaving
the rest half
$$\epsilon_g = \frac{h^2}{8md_c^3}(\Delta d).  \eqno(44) $$

\noindent
as the net fall in the energy of an electron in the locally
strained lattice.  In what follows from a detailed analysis
[53] of certain simple representative examples of trapped
quantum particle(s) interacting with oscillating particle(s),
we also find that $q$ of such an electron oscillates with the
frequencies of lattice oscillations ({\it i.e.} phonons). To
understand the this inference, without going through the dertails
available in [53], it may be noted that zero-point energy
($\varepsilon_o$) and momentum ($q_o$) of a particle constrained
to move through some sort of channels in the lattice structure,
obviously, depends on the diameter/ width ($d_c$) of the channel.
Naturally, when $d_c$ oscillates with a lattice oscillation
at a phonon frequency, $\varepsilon_o$ and $q_o$ also oscillate
with the same frequency and
the electron and lattice can be seen to exchange energy/ momentum
from each other [53].  Since an electron remains in this state
unless it receives $\epsilon_g$ energy from outside,
$\epsilon_g$ can be identified as an {\it energy gap} between its
state with strained lattice and that with zero-strained lattice.
Further as each electron in our theoretical framework represents
({\bf q}, -{\bf q}) pair, the existence of this gap means that
conduction electrons in the fluid are in a state of
({\bf q}, -{\bf q}) bound pairs and the effective free energy
of $q-$motions can be expressed by
$$N\varepsilon_o' = N\varepsilon_o - N\epsilon_g(T) =
N\varepsilon_o - E_g(T)    \eqno(45) $$

\noindent
where $E_g(T)$ is the net decrease in the free energy of the
electron fluid.  Since as discussed in Section 2.0, each electron
binds with lattice and $N-1$ other conduction electrons, $E_g(T)$
could be identified as the collective binding of all electrons
in the solid.  However, it is argued (Section 6.4 below), that
this gap does not show its effectiveness unless the system is
cooled to $T < T_c (\equiv \epsilon_g)$.

\bigskip
\noindent
{\bf 6.4 Transition temperature }

\bigskip
Two electrons (as per Pauli exclusion) can have : (i) different
$K$ and equal $q$ or (ii) equal $K$ and different $q$.  As a
result of the latter possibility many electrons can have $q >
q_o$ at the cost of their $K-$motion energy and the state of
$q = q_o$ even for $T \le T_o$ (excluding $T \le T_c$) does not
attain stability.  Consequently, inter-electron binding induced
by lattice strain does not operate effectively unless the system
is cooled to $T \le T_c (\equiv \epsilon_g)$.  This renders
$$T_c = \frac{h^2}{8\pi mk_Bd_c^2}\frac{\Delta d}{d_c}=
T_o\frac{\Delta d}{d_c} = T_o\frac{\beta l}{d_c}  \eqno(46)$$

\noindent
with $T_o =h^2/(8\pi mk_Bd_c^2)$, and $l = a$ (representing the
{\it interatomic separation} in conventional superconductors) or
$l = c$ ({\it the lattice parameter perpendicular to the
conduction plane of electrons in HTC systems}).  In Eqn. 46, we
use $\Delta d = \beta l$ in view of the fact that the strain
$\Delta d$ should be proportional to $l$ with proportionality
constant $\beta$ representing a kind of the elastic property of
interatomic bonds (in conventional systems) or the lattice
parameter $c$ in HTC systems.  It is evident that $T_c$ represents
a transition temperature below which the conduction electrons are
in a stable state of their $q-$motions.  Since these electrons
cease to have relative motion (Section 4.2), they move in order of
their positions without any collision or scattering which means
that they only have correlated motion without disturbing their
relative positions in real and phase spaces.  The stability of
this state is not disturbed by any perturbation of energy
$< \epsilon_g$, ({\it viz.} external magnetic field, electric
current {\it etc.}), which indicates that {\it the long range
electron-electron correlations
mediated by phonons} and related properties (superconductivity,
coherence, persistence of currents, {\it etc.}) are also not
disturbed unless the energy of these perturbations crosses its
critical magnitude. The observation of critical magnetic field(s),
critical currents, {\it etc.} support this observation.

\bigskip
\noindent
{\bf 6.5  Nature of transition}

\bigskip
The well known fact that $A'$ does not become critical
at any $T$ [51] implies that it has no energy change
at $T_c$.  On the other hand $N\varepsilon_o$ has only
marginal change with the process of straining lattice
(or the formation of ({\bf q}, -{\bf q}) bound pairs) which
starts at $T_a$ and continues till $T = 0$.  Evidently,
the changes from $N\varepsilon_o(T_a)$ to
$N\varepsilon_o(T = 0) = N\varepsilon_o(T_a) - E_g(T = 0)$
lasts over a wide range (from $T_a$ to $T = 0$) of $T$.
To this effect it may be noted that electrons assume their
bound pair states at $T \approx T_a > T_c$.  Naturally, the
fall in their energy by $E_g(T = 0)$ has already taken
place but because of Pauli exclusion $E_g(T = 0)$
does not assume its effectiveness unless the system is
cooled through $T_c$.  This clearly means that
$N\varepsilon_o$ passes smoothly from
$N\varepsilon_o(T_c^+)$ to $N\varepsilon_o(T_c^-)$,
{\it i.e.} without any jump in energy.  Evidently, the
transformation of the electron fluid into its superconducting
state at $T_c$ is a {\it second order transition}.

\bigskip
When electrons at $T_c$ move from their state
of $\lambda/2 < d_c$ (normal fluid state at $T = T_c^+$)
to that of $\lambda/2 = d_c$ (superconducting state at
$T = T_c^-$), their relative
$\phi-$positions ($\phi = kd$) change from  $\phi > 2\pi$
to $\phi = 2\pi$.  This means that electrons move from their
configuration of random locations in $\phi-$space to that
of orderly separated $\phi-$positions ($\Delta\phi
= 2n\pi$ with $n = 1,2,3, ..$).  In other words the electrons
have a kind of order-disorder transformation in $\phi-$space
with their transformation to superconducting state at $T_c$.
Evidently, because of this order, electrons in superconducting
state maintain a definite phase correlation and exhibit
coherence of their motion as well as quantized vortices.

\bigskip
\noindent
{\bf 6.6 Typical estimates of $T_c$}

\bigskip
The universal component of the hamiltonian $H_o(N)$ (Eqn. 2)
of electron fluid in a solid does not differ from
that of liquid $^3He$ (if spin-spin interaction and spin-orbital
interactions are excluded in this case too).  Evidently,
superfluid $T_c$ for
both fluids can be obtained by Eqn. 46.  Since the desired
experimental data about $d$ and strain $\Delta d$ of reasonably
high accuracy are available for liquid $^3He$, it is instructive
to determine its $T_c$ from Eqn. 46 and compare the same with
experimental value to have an idea about the accuracy
of Eqn. 46.  Therefore, we use the density data available from
[52] to determine (i) $d = 3.935718 \AA$ at $T = 0.6$K at which
the volume expansion (or onset of $He-He$ bond strain is
observed), (ii) $d = 3.939336 \AA$ at $T = 0.1$K and (iii)
$\Delta d = 0.003618 \AA$ to find $T_c = 1.497$mK which agrees
very closely with experimental $T_c \approx 1.0 $mK [54, 55].
It may be observed that no other theory [56] has predicted a
$T_c$ for liquid
$^3He$ which falls so close to the experimental value.
Evidently, this indicates the accuracy of Eqn. 46.

\bigskip
Although, crystal structural data for widely different
superconducting solids are available in the literature,
and one can use these data to determine the inter-particle
distance but there is no way to find an accurate value of the
channel size $d_c$ through which conduction electrons flow and
strain $\Delta d_c$ produced by $f_o$ of electrons.
Consequently, one can use Eqn. 46 for electron fluid only to
estimate the range of typical values of $T_c$ by using
typical numbers for $d_c$ and $\Delta d_c$.  To this effect we
first find that the force constant $C_o = 2.735$ dyne/cm
(estimated from $C_o = 3h^2/4md^4$) related to $f_o$ for
liquid $^3He$ matches closely with $He-He$ single bond force
constant $\approx 2.0$ dyne/cm estimated from zero wave vector
phonon velocity 182 m/sec [52].   A similar estimate of $C_o$
for the $f_o$ of electrons can be made by using
(i) $d_c = 3.935718 \AA$ ({\it i.e.} as large as $d_{He-He}$)
and (ii) as short as $d_c = 1.0 \AA$ which is expected to
represent the typical $d_c$ for superconducting solids.  Using
standard value of electron mass $m_e = 0.9109$x$10^{-27}$ gm,
we, respectively, found $C_o = 15$x$10^{3}$ dyne/cm and
$C_o = 36.0$x$10^{5}$ dyne/cm which compares well with the
typical force constants for a bond between two nearest
neighbors in widely different solids.
In view of this observation, we
assume that the strain factor $\Delta d/d$ in superconducting
solids approximately has the same value (= 9.1897x$10^{-4}$)
that we observe experimentally for liquid $^3He$ and use
Eqn. 46 to find $T_c = $ 8.23 K for $d_c = 3.935718 \AA$
and $T_c = $124 K for $d_c = 1.0 \AA$ which closely fall in
the range of experimentally observed $T_c$.  Evidently,
Eqn. 46 explains the experimentally observed $T_c$ for
conventional superconductors as well as HTC systems.

\bigskip
\noindent
{\bf 6.7 Factors affecting $T_c$}

\bigskip
Since conduction electrons in a solid move in an interacting
environment, $m$ in Eqn. 46 could be replaced by $m^*$ (the
effective mass of the electron).  Evidently, $T_c$ depends
on channel size $d_c$, strain $\beta l$, and $m^*$ which means
that one may, in principle, change $T_c$ at will if there is
a method by which these parameters for a given solid can be
suitably manipulated.  However, any controlled change in these
parameters does not seem to be simple.  For example we may apply
pressure to decrease $d_c$ in order to increase $T_c$ but the
compression produced by pressure may increase electron-lattice
interactions in such a way that an increase in $m^*$ may
overcompensate the expected increase in $T_c$ and one may
find that $T_c$ decreases with increase in pressure.  Evidently,
though $T_c$ is normally expected to increase with pressure,
its pressure dependence, for some superconductors, may show
opposite trend or complex nature.  Similarly, we can take the
example of a change in $T_c$ with $\Delta d$ which equals
$\beta c$ for a HTC system and $\beta a$ for a conventional
superconductor.  Since $\beta c$
is much larger than $\beta a$, lattice strain could be one
factor which may increase $T_c$ of a HTC system by a factor of
$c/a$, if $d_c$, $\beta$, $m^*$, {\it etc.} for two types of
systems do not differ.  As analyzed by Leggett [57], $T_c$
increases with the number of conducting planes ($n_{cp}$) per
unit cell for certain groups of HTC systems which indicates that
$T_c$ realy increases with $c$, since $c$ increases with $n_{cp}$.
However, $T_c$ does not increase with $n_{cp}$ always [57] and
we find that the dependence of $T_c$ on $d_c$, $\beta$, $m^*$
and $\Delta d$ is not so simple as it appears from Eqn. 46.
One may hope that there could be some possible mechanism
which may help in raising $T_c$ by manipulating these parameters.
In this context, it may be emphasized that Eqn. 46 does not
rule out the possibility of achieving {\it room temperature} (RT)
superconductivity since raising $T_c$ from 124K to 300K simply
requires a system where $(1/m^*d_c^2)(\Delta d/d_c)$ is increased
from 1.0 to 2.5 which can achieved if $m^*$ alone changes from
$m$ to $0.4m$ or $\Delta d/d_c$ changes from $0.001$ to $0.0025$
or $d_c$ is reduced by a factor of 1.6.  As a matter of principle
any change or perturbation, which adds ({\it removes}) energy
to ({\it from}) the $q-$motions, will decrease ({\it increase})
$T_c$.

\bigskip
\noindent
{\bf 6.8 Strain energy of lattice}

\bigskip
We note that the strain in lattice produced by ({\it say}) $i-$th 
electron is a local effect and its magnitude depends on the
quantum size ($\lambda_i/2$) and hence the momentum ($q_i$) of the 
electron which renders $\epsilon_s = \epsilon_s(q_i)$.  However, 
since identical local strains are produced by all conduction
electrons distributed uniformly in the solid, a collective long
range impact of these strains can be observed due to strong
inter-atomic forces,
and the net strain energy of the lattice can be expressed as
$E_s = E_s(q_1, q_2, q_3, ...)$.  Naturally, because of this
dependence of $E_s$ on momenta of all conduction electrons,
a sustained exchange of energy between electrons and lattice
is expected when the channel size oscillating with a lattice
oscillation leads the quantum size of different electrons
to oscillate with the same frequency.  

\bigskip
Considering a simple unit of two electrons (say e1 and e2)
separated by a small
lattice block between them, it can be easily visualized that two 
electrons will gain ({\it lose}) energy from the strained lattice
if the said lattice block has a kind of breathing oscillation with 
an expansion ({\it contraction}) in its size leading to a decrease
({\it increase}) in the strain and the strain energy of the said
block; this will also render a decrease ({\it increase}) in the
size of two channels occupied by e1 and e2 for which their
$\varepsilon_o$ would obviously increase ({\it decrease}). However,
if the position of the said block oscillates around its CM without
any change its size, e1 and e2 would exchange energy with each
other. If the block moves to wards e1, it decreases $d_c$ for e1
and increases $d_c$ for e2, and in the process $\varepsilon_o$(e1)
increases at the cost of $\varepsilon_o$(e2) and the necessary
energy flows through appropriate mode(s) of phonons from e1 to e2
and vice versa. It is evident that the dynamical motions in a
solid are complex and the two motions that we considered in this
example could be the simplest possible modes but these examples
explain the typical nature of the process of energy exchange
between two electrons and lattice through phonons. Since $E_s$
stays with the lattice even at $T=0$ at which no phonon exists,
$E_s$ serves as the source of phonons and supports phonon
mediated correlated motion of conduction electrons by sustained
energy exchange between electrons and lattice at all $T \le
T_c$ including $T = 0$.

\bigskip
\noindent
{\bf 6.9 Order parameter(s)}

\bigskip
The conduction electrons in their superconducting state
are in the ground state of their $q-$motions with free
energy $N\varepsilon_o - E_g(T)$.  Since $N\varepsilon_o$
is a constant value
which depends on the relevant parameters of the solid
at $T = T_a$, only $E_g(T)$ is crucial for different
aspects of superconducting state.  The lattice
strain, on which $E_g(T)$ depends, can be identified as
the basic order parameter of the transition.  However,
since the conduction electrons below $T_c$ assume a
configuration where they have : (i) some sort of
localization in their positions in the real space unless
they are set to move, (ii) an ordered structure in
$\phi-$space defined by $\Delta\phi = 2n\pi$ with
$n = 1,2,3, ...$, (iii) definite momentum $q = q_o'$,
(iv) definite orientation of their spins {\it if different
interactions involving spins so prefer} ({\it cf.}
Section 6.11), (v) definite amount of superfluid density
$\rho_s$ ({\it cf.} Sections 7.2 and 7.3),
etc.  Naturally, electrons at $T \approx T_c$ must have
large amplitude position fluctuation (leading to charge
density fluctuation), $\phi-$fluctuation, momentum
fluctuation, spin fluctuation, $\rho_s-$fluctuation,
{\it etc.} which should naturally couple with the lattice
strain identified as the basic order parameter of the
superconducting transition.  However, the nature of coupling
may differ from system to system.

\bigskip
\noindent
{\bf 6.10 Comparison with normal state}

\bigskip
Although, each conduction electron in its macro-orbital picture 
represents a pair of electrons moving with {\bf q} and -{\bf q}
momenta in their CM frame which moves with momentum {\bf K} in
the laboratory frame, however, as discussed in Section 6.4, Pauli
exclusion does not allow this pair to assume stability at $T \ge
T_c$.  Consequently, like the particles of any normal liquid,
electrons in their normal fluid state have random motions and
inter-electron and electron-lattice collisions.  However, with the
onset of mechanical strain in the lattice structure at $T_c$,
electrons assume the configuration of ({\bf q}, -{\bf q}) bound
pairs with well defined binding energy ($\epsilon_g$) and $q =
q_o'$ and as discussed in Section 6.3 they have sustained
energy exchange with lattice.  Similarly they also have sustained
$\phi-$correlation ({\it cf.} Section 6.5) with well defined
positions ($\phi = kr = 2qr = 2n\pi$) in $\phi-$space which
means that they have a kind of orderly arrangement in real space
since they all have $q = q_o'$.  In other words the conduction
electrons in their superconducting state have sustained $q-$,
$r-$ and $\phi-$correlations which are not seen in their normal
state.  Further as the conduction electrons in their normal
state exchange energy with the lattice by way of their collisions
with the lattice constituents but such collisions do not exist in
the ground state ({\it cf.} Section 4.2) that exhibits
superconductivity.

\bigskip
In view of our inferences made in Section 4.2, conduction
electrons can move only in order of their locations in the
conduction channel, in case they are set to move.  This orderly
motion of electrons with their definite $\phi-$separation ($\phi
= 2n\pi$) clearly represents a kind of coherence in their motion
exhibited by their superconducting state.  Further since
$\phi = kr = 2n\pi$ implies that $k$ and $r$ of two electrons in
superconducting state are inter-dependent, their binding in
momentum space, obviously, represents a binding in real space.
They have no kinetic energy above their zero-point energy of
their localization in the channel since their $K-$motions get
delinked from $q-$motions ({\it cf.} Section 7.2).
Consequently, under the influence of their mutual
repulsion and their interaction with other lattice
constituents, their positions in the real space are expected
to define a crystalline arrangement.  Naturally, their mutual
repulsion can also be an important factor to facilitate their
motion in the order of their positions in a channel and this
fact helps in identifying the basic difference of their
superconducting state with their normal state where electrons
are free to move randomly.  One may identify this difference
with the difference in the orderly movement of parading soldiers
of an army platoon and the movement of people in a crowd.

\bigskip
\noindent
{\bf 6.11 Co-existence with other properties}

\bigskip
Since the conduction electrons in their superconducting
state get localized with an orderly arrangement in real
space like atoms in a crystal ({\it cf.} Section 6.10)
({\it of course with a freedom to move in order of their
positions}), they cease to have inter-electron collisions
as well as collisions with lattice.  Consequently, their
spins can sustain their definite orientations leading
to a magnetic state, such as diamagnetic or ferro-magnetic
or anti-ferromagnetic state.  The nature of the magnetic state
is, obviously, decided by the different interactions with
electrons spins and it can be understood by using the
well known theories of diamagnetism, or ferro-magnetism
or anti-ferromagnetism of a solid.  Evidently, our
theoretical framework finds no compelling reason for the
superconducting state to be diamagnetic only.  In fact the
magnetic state of the
superconducting electron fluid in a particular solid
should be governed by the condition of minimum free
energy with respect to an appropriate order-parameter.
Both, the diamagnetism of most superconductors and the 
co-existence of superconductivity with ferro-magnetism
in fewer systems, could be a simple consequence of
this condition.  For the similar reasons, we may argue
that pairing of electrons can also occur in triplet $p-$state
or singlet $d-$state.  

\bigskip
\noindent
{\bf 6.12 Principles of Superconductivity }

\bigskip
Recently, Mourachkine [58] analyzed general principles of
superconductivity from the standpoint of practical realization
of RT superconductivity.  He observes that : (i) RT
superconductivity, if ever realized, would not be BCS type,
(ii) the quasi-particle pairing which takes place in momentum
space could possibly take place in real space and if it happens
BCS theory and future theory of unconventional superconductors
can hardly be unified, (iii) the mechanism of electron pair
formation in all superconductors differs from the mechanism of
Cooper pair condensation, (iv) the process of electron pairing
precedes the process of Cooper pair condensation, {\it etc.}
In this context
our theoretical analysis reveals the following: (a) The
main factor, which induces an indirect attraction between two
conduction electrons necessary for the
formation of Cooper type pairs, is a kind of mechanical strain
in the lattice produced by the zero-point force of conduction
electrons; while this fact supplements the BCS model in certain
respect but at the same time it underlines the fact that the
real mechanism of pairing of electrons responsible for
superconductivity of widely different solids differs from BCS
theory, (b) the quasi-particle electron pairing takes
place not only in momentum space as envisaged by BCS model but
in certain sense it, indirectly, occurs also in $r-$space
({\it cf.} Section 6.10) as well as in
$\phi-$space, (c) While the
conditions, in which electron pair formation is possible, exist
at $T \le T_o$ but the stabilization of such pairs (represented
by an energy gap), which goes hand in hand with the onset of
superconductivity or Cooper pair condensation
occurs at $T_c$ (orders of magnitude lower than $T_o$); this
clearly shows that the process of binding and the process of
pair condensation are different and the former
precedes the latter.  As such these facts indicate that our
inferences agree to a good extent with the basic principles of
superconductivity as envisaged by Mourachkine [58].  However,
in variance with his observation, BCS type model with
mechanical strain produced by zero-point force as the
main origin for the phonon induced interaction leading
to Cooper type bound pairs of electrons unifies our
understanding of the superconductivity of widely different
superconductors.  

\bigskip
\centerline {\bf 7.0 Consistency With Other Important Theories}

\bigskip

\noindent
{\bf 7.1 BCS theory}

\bigskip
On the one hand our theoretical framework reinforces the basics
of BCS picture ({\it viz.}, the formation of Cooper type pairs
of electrons and their condensation as the origin of
superconductivity), on the other hand it differs in certain
respect.  For example, it identifies the mechanical strain
in the lattice produced by the zero-point force of conduction
electrons (Sections 3.4.5 and 6.2) as the main factor responsible
for the phonon induced attraction between two electrons.  It is
evident that the electrical strain emphasized by BCS model too
contributes to this attraction but it may noted that the
mechanical strain alone predicts a $T_c \approx 124$K
({\it cf.} Section 6.6), while the electrical strain in BCS
picture accounts for a $T_c <$ $\approx$ 25K only.  Assuming that
both strains contribute in all systems and these two values
represent their proportional contributions, one finds that
electrical strain contributes only around $18 \%$.  Further
it is important to note that the lattice in the superconducting
phase stores an additional potential energy as the strain
energy, $E_s$ ({\it cf.} Section 6.8), when the energy of
net system (conduction electrons + lattice) falls with the
onset of Cooper type pair formation.  Since $E_s$ stays with the
lattice even at $T=0$ at which no phonon exists in the system,
this energy becomes a source for the creation of phonons
necessary to mediate correlated motion of two electrons of a
Cooper type pair at all $T \le T_c$ including $T=0$.  As such
this section not only identifies the basic differences of
our theory with BCS theory but also underlines the fact
that our theory incorporates BCS theory.  Different aspects
of superconducting phase such as coherence length, ctrical
current, critical magnetic field, persistence of current,
{\it etc.} which depend on $E_g(T)$ can, therefore, be
understood by using the appropriate relations available
from BCS theory. 

\bigskip
\noindent
{\bf 7.2 Two fluid theory}

\bigskip
We note that: (i) each electron represented by a macro-orbital
has two motions, $q$ and $K$, (ii) they have separate free
energy contributions, $N\varepsilon_o$ and $A'$ and (iii)
the onset of superconductivity locks the $q-$motions of all
electrons at $q = q_o$ with an energy gap which isolate them
from $K-$motions.  Evidently, the superconducting state of
the fluid at $T \le T_c$ can be described by
$$\Psi^S(N) = \Pi_i^N\zeta_{q_o}(r_i)
\sum_P^{N!} (\pm 1)^P\Pi_i^N\exp{[i(P{\bf K}_i{\bf R}_i)}]
\eqno(47)$$

\noindent
which has been obtained by using all $q_i = q_o$ in
Eqn. 28.  We note that $\Psi^S(N)$ is product of two separate
functions, {\it i.e.}, $\Psi^S(N) = \psi_K(N)\psi_q(N)$ with 
$$\psi_K(N) = \sum_P^{N!}(\pm 1)^P
\Pi_i^N\exp{[i(P{\bf K}_i{\bf R}_i)}] \eqno(48)$$

\noindent
and
$$\psi_q(N) = \Pi_i^N\zeta_{q_o}(r_i) \eqno(49)$$

\noindent
This implies that the electron fluid at $T \le T_c$ can be
identified as a homogeneous mixture of two fluids: (F1)
described by $\psi_K(N)$ where electrons represent
some sort of quasi-particles described by plane waves
of momentum $K$ and (F2) described by $\psi_q(N)$ where
each electron represents a kind of localized particle
in ({\bf q}, -{\bf q}) bound pair state where it ceases
to have collisional motion.  With all electrons having
$q = q_o$, F2 represents the $q-$motion ground state where
each electron has no thermal energy ({\it i.e.}, no energy
above the zero-point energy, $\varepsilon_o$) and it
has {\it zero entropy} which is also supported by the
fact that the number of different configurations with
all particles having $q = q_o$ is only 1.  We further
note that particles in F2 are
basically localized; if they are set to move they move in
order of their location with no relative motion, no collision
or scattering.  Naturally, they find no reason
(such as collision or scattering) to lose their energy
indicating that their flow should be resistance free implying
that F2 is in superconducting phase.   Since each particle in
this phase has an energy gap ($\epsilon_g$) with respect
to its state in normal phase at $T_c^+$ (just above $T_c$),
the former is stable against any perturbation of energy
$< \epsilon_g$. Naturally, when this fact is clubbed with
the coherent motion of macroscopically large number of
electrons it becomes evident that the source resistance
should be strong enough to reduce the velocity all such
electrons in a single event which however
is not possible.  As such we find that F1 and F2 at all
$T \le T_c$ have all properties that have been envisaged by
Landau [59] in the normal fluid and superfluid components of an
electron fluid which implies that our theory provides
microscopic foundations for the two fluid phenomenology.
Since Bardeen [60] has elegantly analyzed BCS theory
as the microscopic basis of two fluid theory and our theory
incorporates BCS model, his results can be used
to determine the normal ($\rho_n$) and superfluid density
($\rho_s$) components needed for implementation of two
fluid theory.

\bigskip
\noindent
{\bf 7.3 $\Psi-$ Theory }

\bigskip
We note that superconductivity is basically a property of
F2 in its ground state which represents the relative
configuration of electrons in ({\bf q} -{\bf q}) pair
states with $q = q_o$ which is, evidently, described by
$$\Psi_o(N) = \Pi_i^N \zeta_{q_o}(r_i) = \sqrt{\rm n} \eqno (50)$$

\noindent
(with ${\rm n} = N/$V being the electron number density).
To obtain Eqn. 50 we separated the $K-$dependent part of
$\Psi^S(N)$ (Eqn. 47) describing F1.  Since each electron in
$({\bf q} -{\bf q})$ configuration under the influence of any
perturbation that makes it move with a momentum say
$\Delta {\bf K}$ assumes ({\bf q} + $\Delta${\bf K},
-{\bf q} + $\Delta${\bf K}) configuration and its state is
described by
$$\zeta{(r, R)} = \zeta_{q_o}(r)\exp{(i{\bf Q}.{\bf R})}
\eqno(51)$$

\noindent
with ${\bf Q} = 2\Delta {\bf K}$, it is evident that
superconducting state under such perturbation would be
described by
$$\Psi'_o(N) = \Pi_i^N \zeta_{q_o}(r_i)\exp{(i\Phi)} =
\sqrt{\rm n}\exp{(i\Phi)}  \eqno (52)$$

\noindent
with its phase $\Phi = \sum_i^N {\bf Q}_i.{\bf R}_i$ and
${\bf Q}_i = 2\Delta {\bf K}_i$.  However, for the
phenomenological reasons ({\it viz.} the number density of
superconducting electrons (${\rm n}_s$) need not be equal
to ${\rm n}$) we replace $\Phi$ by $\Phi + i\Phi'$ and
recast $\Psi'_o(N)$ as
$$\Psi'_o(N) = \sqrt{n_s}\exp{(i\Phi)}  \eqno (53)$$

\noindent
which renders ${\rm n}_s = {\rm n}\exp{(-2\Phi'}$).  We note
$\Psi'_o(N)$ clearly has the structure of $\Psi-$function
that forms the basis of the well known $\Psi-$theory of
superfluidity.  This shows that our theory provides microscopic
foundation to the highly successful $\Psi-$theory [20].

\bigskip
\noindent
{\bf 7.4 Theory based on the proximity of a QPT}

\bigskip
In view of Sections 7.2 and 7.3, superconductivity is a property
of F2 system in its ground state obviously representing a state
at $T = 0$K.  This implies that superconducting transition is,
basically, a {\it quantum phase transition} that occurs in F2
system exactly at $T = 0$ but its proximity with F1 makes it
appear at non-zero $T$ in the real system (a homogeneous mixture
of F1 and F2 where each particle participates identically); they
manifest as two separated fluids at all $T \le T_c$ for the
presence of the
energy gap, while such separation ceases to exist at $T > T_c$.
Evidently, our theory is consistent with the idea which relates
superconductivity with the proximity effect of a quantum phase
transition [15].

\bigskip
\centerline {\bf 8.0 Concluding Remarks}

\bigskip
This paper uses a new approach to lay the basic foundations
of superconductivity.  It finds that each conduction
electron (particularly in low energy states) is more accurately
represented by a macro-orbital ({\it cf.}, Section 3.4.7) rather
than a plane wave.  While it reinforces the basics of the
BCS model of superconductivity  ({\it viz.}, the formation
of $({\bf q}, -{\bf q})$ bound pairs of conduction electrons
and their condensation) but it also reveals that
the basic reason for the phonon induced attraction
between two electrons (responsible for the formation of Cooper
type bound pairs) rests with a mechanical strain in the lattice
produced by zero-point force (a well known force of a spatially
confined quantum particle in its ground state).  The electrical
strain in the lattice produced by the electric charge of
electrons can have its contribution to such attraction and
resulting binding of two electrons as Cooper type pair.  In
principle, other interactions such as spin-spin, spin-latice,
{\it etc.} too can add to the said binding which, obviously,
implies that our approach accomodates all possible interactions
that may contribute to bound pair formation.

\bigskip
In agreement with the BCS model, our approach also finds 
an energy gap (between superconducting and normal states
of the electron fluid) resulting from the phonon induced
attraction between two electrons in ({\bf q}, -{\bf q})
configuration.  Evidently, our framework can identically
account for all aspects of superconducting phase that are
accounted for by BCS theory and for this reason we need not
rederive the relations which connect $E_g(T)$ with various
properties of a superconductor and restate the scientific
arguments which help in their physical understanding.  It is,
however, important to note that our theory concludes a simple
relation for $T_c$ ({\it cf.}, Eqn. 46) which not only
accounts for the highest $T_c$ that we know to-day but
also reveals a possibility of observing superconductivity
at RT.

\bigskip
The diameter/ width of the narrow channels ($d_c$)
through which conduction electrons flow in a solid plays an
important role in controlling superconducting $T_c$ (Eqn. 46),
while the process through which conduction electrons come into
existence at a $T \ge T_c$ is unimportant.  Our approach is
also applicable to the systems with holes as the charge
carriers because the flow of holes is nothing but the flow of
electrons ({\it once again through the narrow channels}) by 
way of hopping between successive electron vacancies.

\bigskip
We find that superconductivity is basically a property of the
ground state of electrons as $({\bf q}, -{\bf q})$ bound pairs
with $q = q_o' = \pi/d_c'$.  The system specific or class
specific properties can be obtained by using appropriate
term(s) from $V'(N)$ as perturbation on the states of $H_o(N)$
(Eqns. 1 and 2).  As such our approach does not forbid: (i) pair
formation in triplet $p-$state and singlet $d-$state as well as
(ii) the coexistence of superconductivity with ferro-magnetism or
anti-ferromagnetism.

\bigskip
Although, this paper does not analyze the origin of experimentally
observed pseudo-gap, charge stripes, {\it etc.} in HTC systems,
however, it appears that these aspects do not have {\it direct}
relation with the origin of superconductivity.  We would like to
examine the physics behind these observations as part of our
future course of studies.
   
\bigskip
Finally, it may be mentioned that our theory only assumes that,
to a good approximation, conduction electrons can be identified as
HC particles of a Fermi fluid which flows through narrow channels
(cylindrical tubes or 2-D slots in the lattice structure).  It
makes no presumption about the nature of the microscopic mechanism
of superconductivity.  Its all inferences are drawn from a
systematic analysis of the solutions of the Schr\"{o}dinger
equation of $H_o(N)$.  The fact, that a system described by
$H_o(N)$ exhibits superfluidity if its particles have inherent or
induced inter-particle attraction
and the system remain fluid at $T \le T_c$, has been successfully
demonstrated in [42] by using the same approach.  The mathematical
foundation and formulation of our theoretical framework are simple
and it has great potential for developing equally simple
understanding of different aspects of superconductivity and
related behavior of widely different superconductors.

\bigskip
Over the last two decades, one of the major thrusts of
researches in the field of superconductivity has been
to find the basic mechanism which can account for the
experimentally observed high $T_c$.  One may find that
the prersent work has been able to achieve this objective.
Since it provides clear picture of the ground state
configuration of electrons, it may help in studying
the details of inter-particle correlations at $T \le T_c$
in $q-$, $\phi-$ and $r-$spaces required for understanding
their transport properties in superconducting phase;
however, such studies could be taken only in a future
course of our research.
As evident from Sections 5.0 and 6.0, the present study
also reveals that electron fluid in solids should behave
almost like: (i) a system of non-interacting fermions
at $T > T_a$ when electrons can be represented by plane
waves, (ii) a Landau-Fermi liquid (with quasi-particle
mass = $4m$ which may, however, be modified due to
interacting environment of the electrons) at
$T_a > T > T_c$ when they are better
represented by macro-orbitals, and (iii) a singular Fermi
liquid at $T \le T_c$ when $N\varepsilon_o$ becomes
critical under the influence of zero-point force.  Varma
{\it et.al.} [56] have elegantly itroduced the subject
related to these three phases of the behavior of a system
of interacting fermions. 

\bigskip
The present theory is, obviously, applicable to any other
system of HC fermions with weak inter-particle attraction
({\it viz.} liquid $^3He$) by simply assigning the role of
{\it lattice structure} to the atomic arrangement of
neighboring $^3He$ atoms around a chosen $^3He$ atom whose
$q_o$ is now decided by $d = (V/N)^{1/3}$.  In this context,
it may be mentioned that no other theory has been able to
obtain superfluid $T_c$ for liquid $^3He$ which falls so close
to its experimental value as found by us ({\it cf.} Section 6.6).
Although our account of this system is not expected to differ
from that found by using BCS model because our theory
incorporates BCS model but for the first time it identifies
zero-point force leading to a strain in $He-He$ bonds as the
origin of Cooper type bound pairs of $^3He$ atoms responsible
for the superfluidity of liquid $^3He$.  While the salient
aspects of our study of liquid $^3He$ (presented in a recent
conference) are available in [42], the detailed would soon
be published elsewhere.

\bigskip
\noindent
{\it Note :} The author would greatly appreciate the
comments of the scientific community on this simple
approach to the understanding of superconductivity of
widely different systems.

\newpage

\centerline{\bf Appendix - A}

\bigskip
\centerline{\bf A Critical Analysis of $<A\delta{(x)}>$}

\bigskip
The relative motion of two particles interacting through a
central force potential basically represents a 1-D motion along
the line joining their centers of mass.  The following analysis
uses this observation to establish the validity of
$<V_{HC}(r)> = <A\delta{(r)}> = 0$ (Eqn. 18) for all possible
physical situations in relation to
the relative motion of two HC particles in 1-, 2- and
3-dimensions by considering the 1-D analogue ($<V_{HC}(x)> =
<A\delta{(x)}> = $) of Eqn. 18.  We note $A$ in $V_{HC}(x)
\equiv A\delta{(x)}$ is such that $A \to \infty$ for $x \to 0$
and it can in general be expressed as        
$$A = Bx^{-(1+ \alpha)}    \eqno(A-1)$$

\noindent
where both $B$ and $\alpha$ are $ > 0$.  Using the pair
state $\zeta^-$ or $\zeta^+$ (Eqns. 13 and 14),
we find that
$$<A\delta{(x)}> =
B\frac{2\sin^2{(kx/2)}}{x^{(1+ \alpha)}}|_{x=0} \eqno(A-2)$$

\noindent
is an in-determinant which can be simplified to
$Bk^2x^{1-\alpha}/2$ for $x \approx 0$. Evidently,
when $x \to 0$, $<A\delta{(x)}>$ has $0$ value for
$\alpha < 1$, a $+ve$ value (= $Bk^2/2$) for
$\alpha = 1$ and $\infty$ for $\alpha > 1$.  Since no
physical system can ever occupy a state of $\infty$
potential energy, $\alpha > 1$ corresponds to a
physically uninteresting case.  While remaining
$\alpha$ values correspond to physically possible
configurations, $\alpha = 1$ is the sole
point on the $\alpha-$line for which $<A\delta{(x)}>$
assumes a finite $+ve$ value.  In fact $\alpha = 1$
stands as a sharp divide between the states
of $<A\delta{(x)}> = 0$ and $<A\delta{(x)}> = \infty$.
To understand the physical significance of these
results, we note the following.

\bigskip
\noindent
1. $<A\delta{(x)}> = 0$ for $\alpha < 1$ implies that 
Eqn. 18 is clearly valid for this range of $\alpha$.

\bigskip
\noindent
2. $<A\delta{(x)}> = Bk^2/2$ for $\alpha = 1$
renders
$$E^* = \frac{\hbar^2k^2}{4m} + \frac{Bk^2}{2} =
\frac{\hbar^2k^2}{4m}\left(1 + \frac{2Bm}{\hbar^2}\right)
\eqno(A-3)$$

\noindent
which, {\it in principle}, represents the total energy
expectation of the relative motion of two HC particles
interacting through $A\delta{(x)}$.  One may write $E^* =
\hbar^2k^2/4m^*$ to absorb $<A\delta{(x)}> = Bk^2/2$ and 
$\hbar^2k^2/4m$ into a single term by defining $m^*$ as 
$$m^* = \frac{m}{1 + 2Bm/\hbar^2}     \eqno(A-4)$$

\noindent
and use $<A\delta{(x)}> = 0$.  While this shows that
our results, interpretations and conclusions based on
Eqn. 18 are valid even for $\alpha = 1$ if $m$ is
replaced by $m^*$, however,
it does not explain why $E^*$ far from $x=0$ should
be different from $E_k = \hbar^2k^2/4m$ and why $<A\delta{(x)}>$
(as indicated by its proportionality to $k^2$) should
be kinetic in nature; it may be noted that
$<A\delta{(x)}> = Bk^2/2$ does not have potential
energy character of $A\delta{(x)}$ because it is
neither a function of $x$ nor of $<x>$.  Evidently,
$<A\delta{(x)}> = Bk^2/2$ needs an alternative
explanation ({\it cf.} points 3-5 below).

\bigskip
\noindent
3.  Two particles in their relative motion have only
kinetic energy ($E_k = \hbar^2k^2/4m$) till they reach
the point of their collision at $x=0$ where they come
to a halt and $\hbar^2k^2/4m$ gets transformed into an
equal amount of potential energy (as a result of energy
conservation), naturally, proportional to $k^2$ as
really found with $<A\delta{(x)}> = Bk^2/2$.  This mplies
that $<A\delta{(x)}> = Bk^2/2$ does not represent
an additional energy to be added to
$-<(\hbar^2/m)\partial_x^2> = \hbar^2k^2/4m$ in
determining $E^*$ as found in Eqn. A-3.  To this
effect we find that the physical meaning of non-zero
$<A\delta{(x)}>$ of an ill behaved potential function
$A\delta{(x)}$ may differ from that of $<V(x)>$
of a well behaved ({\it i.e.} continuous
and differentiable) potential function, $V(x)$.

\bigskip
\noindent
4.  We also find that $<A\delta{(x)}> = Bk^2/2$ is
independent of the limits of integration $x^-$ and
$x^+$ (with $x=0$ falling between $x^-$ and $x^+$), even
when we use $x^- = -\epsilon$ and $x^+ = +\epsilon$ with
$\epsilon$ being infinitely small.  In other words
$<A\delta{(x)}>$ has solitary contribution (=$Bk^2/2$)
from $x=0$, while $-<(\hbar^2/m)\partial_x^2> =
\hbar^2k^2/4m$ (kinetic energy) has zero contribution
from this point; in fact $-<(\hbar^2/m)\partial_x^2> =
\hbar^2k^2/4m$ is independent of the inclusion or
exclusion of $x=0$ in the related integral.  Evidently,
the energy measured as $-<(\hbar^2/m)\partial_x^2>$
appears as non-zero $<A\delta{(x)}>$ at $x=0$ and $E^*$
should be simply equal to $-<(\hbar^2/m)\partial_x^2>$
by treating non-zero $<A\delta{(x)}>$ as ficitious that
could be assumed to be zero for all practical purposes;
this falls in line with an important observation by
Huang [47] that HC potential is no more than a boundary
condition for the relative wave function.

\bigskip
\noindent
5. In the wave mechanical framework, two colliding
particles either exchange their positions
(across the point $x =0$) or their momenta.  In the
former case they can be seen to cross through their
$\delta-$potential possibly by some kind of
tunneling (in which their kinetic energy does not
transform into potential energy), while in the latter
case they return back on their path after a halt at
$x = 0$ in which case their potential energy rises
at the cost of their kinetic energy.  It appears
that the two possibilities can be, respectively,
identified with $<A\delta{(x)}> = 0$ and
$<A\delta{(x)}> = Bk^2/2$, however, one has
no means to decide whether the two particles
exchanged their positions or their momenta which
implies that the two situations are
indistinguishable and $<A\delta{(x)}>$ can be measured
to have $0$ to $Bk^2/2$ values.  Apparently this is
not surprising since the state of a collision of two
HC particles at $x=0$ ({\it i.e.} an exact $x$) is
a state of zero uncertainty in $x$ and infinitely
high uncertainty in momentum $k$ or in energy
$E_k = \hbar^2k^2/4m$.

\bigskip
In summary non-zero $<A\delta{(x)}> = Bk^2/2$
observed for $\alpha = 1$ should treated as
fictitious.  It can best be attributed to
energy conservation at $x = 0$.  This
implies that $<A\delta{(x)}> = 0$
({\it i.e.} Eqn. 18) is relevant both for
$\alpha < 1$ and $\alpha = 1$.


\newpage

\begin{thebibliography}{??}
\bibitem[1]{Kn:gnus}
J. G. Bednorz and K. A. M\"{u}ller, Z. Phys. B {\bf 64} 169
(1987),
\bibitem[2]{Kn:gnus}
J. Bardeen, L.N. Cooper and J.R. Schrieffer, Phys.
Rev. {\bf 106}, 162 (1957).
\bibitem[3]{Kn:gnus}
E. Dagatto,
Rev. Mod. Phys. {\bf 66}, 763 (1998).
\bibitem[4]{Kn:gnus}
M. A. Kastner, R.J. Birgeneau, G. Shirane, and Y. Endoh,
Rev. Mod. Phys. {\bf 70}, 897 (1998).
\bibitem[5]{Kn:gnus}
M. Imada, A. Fujimori, and Y. Takura, 
Rev. Mod. Phys. {\bf 70}, 1039 (1998).
\bibitem[6]{Kn:gnus}
M.B. Maple,
J. Magn. Magn. Mater. {\bf 177}, 18 (1998).
\bibitem[7]{Kn:gnus}
T. Timusk, and B. Statt,
Rep. Prog. Phys, {\bf 62}, 61 (1999).
\bibitem[8]{Kn:gnus}
J. Orenstein, and A.J. Millis,
Science  {\bf 288}, 468 (2000).
\bibitem[9]{Kn:gnus}
J.L. Tallon, and J.W. Loram,
Physica C {\bf 349}, 53 (2001).
\bibitem[10]{Kn:gnus}
J.C. Campuzano, M.R. Norman, and M. Randeria, in {\it Physics of
concentional  and  unconcentional superconductors}, edited by K.H.
Bennemann and J.B. Ketterson (Springer, New York, 2002)
\bibitem[11]{Kn:gnus}
A. Damascelli, Z. Hussain, and Z.X. Shen,
Rev. Mod. Phys. {\bf 75}, 473 (2003).
\bibitem[12]{Kn:gnus}
G. Deutscher, 
Rev. Mod. Phys. {\bf 77}, 109 (2005)
\bibitem[13]{Kn:gnus}
Z. X. Shen, A. Lanzara, S. Ishihara, and N. Nagaosa,
Philos. Mag. B {\bf 82}, 1349 (2002).
\bibitem[14]{Kn:gnus}
E.W. Carlson, V.J. Emery, S.A. Kivelson and D. Orgad,
in {\it Physics of
concentional  and  unconcentional superconductors}, edited by K.H.
Bennemann and J.B. Ketterson (Springer, New York, 2002).
\bibitem[15]{Kn:gnus}
S. Sachdev
Rev. Mod. Phys. {\bf 75}, 913 (2003).
\bibitem[16]{Kn:gnus}
M.R. Norman, and C. Pepin,
Rep. Prog. Phys. {\bf 66}, 1547 (2003).
\bibitem[17]{Kn:gnus}
V.J. Emery and S.A. Kivelson,
arXiv:cond-mat/9809083; v2 (July 10, 2004)
\bibitem[18]{Kn:gnus}
E. Demler, W. Hanke and S. C. Zhang,
Rev. Mod. Phys. {\bf 76}, 909 (2004)
\bibitem[19]{Kn:gnus}
Y. Yanase, J. Phys. Soc. Japan {\bf 73}, 1000 (2004).
\bibitem[20]{Kn:gnus}
V.L. Ginzburg, 
Rev. Mod. Phys. {\bf 76}, 981 (2004)
\bibitem[21]{Kn:gnus}
A.A. Abrikosov,  Rev. Mod. Phys. {\bf 76}, 975 (2004).
\bibitem[22]{Kn:gnus}
P.W. Anderson, Science {\bf 235} 1196 (1987),
G. Bhaskaran, Z. Zou and P.W. Anderson, Solid
State Commun. {\bf 63} 973 (1987).
\bibitem[23]{Kn:gnus}
F.C. Zhang and T. Rice, Phys. Rev. B {\bf 37}, R3759 (1988).
\bibitem[24]{Kn:gnus}
T.A. Maier, M. Jarrell, T.C. Schulthess, P.R.C. Kent and
J.B. White, arXiv:cond-mat/0504529 (2005)
\bibitem[25]{Kn:gnus}
J.R. Schrieffer, X.G. Wen, and S.C. Zhang,
Phys. Rev. B {\bf 39}, 11663 (1989).
\bibitem[26]{Kn:gnus}     
A. Kampf and J.R. Schrieffer, 
Phys. Rev. B {\bf 42}, 7967 (1989).
\bibitem[27]{Kn:gnus}  
A.J. Millis, H. Monien, and D. Pines
Phys. Rev. B {\bf 42}, 167 (1990).
\bibitem[28]{Kn:gnus}  
N.E. Bickers, D.J. Scalapino, and S.R. White, 
Phys. Rev. Lett. {\bf 62}, 961 (1989).
\bibitem[29]{Kn:gnus}  
P. Monthoux, A. Balatsky, and D. Pines
Phys. Rev. Lett. {\bf 67}, 3448 (1991).
\bibitem[30]{Kn:gnus}  
R.B. Laughlin, Scince {\bf 242}, 525 (1988),  
Phys. Rev. Lett. {\bf 60}, 2677 (1988). 
 \bibitem[31]{Kn:gnus}
A.S. Alexandrov, V.V. Kavanov and N.F. Mott,
Phys. Rev. B. {\bf 53} 2863 (1996), B.K. Chakraverty,
J. Renninger and D. Feinberg,
Phys. Rev. Lett. {\bf 81}, 433 (1998).
\bibitem[32]{Kn:gnus}
A.T. Holmes, D. Jaccard and K. Miyaki,
Phys. Rev. B {\bf 69} 024508 (2004) 11 pages;
D. Aoki et. al., Nature {\bf 413}, 613-616 (2001);
T.R. Kirkpatrik et. al.,
Phys. Rev. Lett. {\bf 87}, 127003 (2001);
C. Pfleiderer et. al., Nature {\bf 412} 58-61 (2001);  
S.S. Saxena et. al.,  Nature {\bf 406} 587-592 (2000);
\bibitem[33]{Kn:gnus}
J. Nagamatsu, N. Nakagawa, T. Muranaka, Y. Zenitani
and J. Akimitsu, Nature {\bf 410}, 63 (2001).
\bibitem[34]{Kn:gnus}
K. Shimizu, et.al., Nature {\bf 419} 597 (2002);
V.V. Struzhkin, et.al., Science {\bf 298} 1213 (2002);
J.-Pierre et.al., Nature {\bf 394}, 453 (1998).
\bibitem[35]{Kn:gnus}
J. E. Hoffman, et. al., Science {\bf 295}, 466 (2002).
\bibitem[36]{Kn:gnus}
M. Lange, et. al., Phys. Rev. Lett. {\bf 90}, 197006 (2001).
\bibitem[37]{Kn:gnus}
H.J. Choi, et. al., Nature {\bf 418}, 758 (2002). 
\bibitem[38]{Kn:gnus}
N.D. Mathur, et. al., Nature {\bf 394}, 39 (1998). 
\bibitem[39]{Kn:gnus}
Y. Maeno, T.M. Rice and M. Sigrist,
Physics To-day {\bf 54} 42(2001).
\bibitem[40]{Kn:gnus}
Y.S. Jain, Macro-orbitals and theory of superconductivity,
presented in CMDAYS-04 (August 25-27, 2004) at Department of
Physics, North-Eastern Hill University, Shillong-793 022,
Meghalaya, India): (A brief discussion included in [42]).
\bibitem[41]{Kn:gnus}  
Y.S. Jain, Cent. Euro. J. Phys. {\bf 2} 709 (2004);
arxiv.org/quant-ph/0603233.
\bibitem[42]{Kn:gnus}
Y.S. Jain, A brief report of the invited talk entitled,
"Unification of the physics of bosonic and fermionic systems"
given at conference CMDAYS-04 (August 25-27, 2004) at
Department of Physics, North-Eastern Hill University,
Shillong-793 022, Meghalaya, India); at Ind. J. Phys.
{\bf 79} 1009 (2005).  ({\it Note} : The factor
$|\sin{({\bf k}.{\bf r})/2}|$
in Eqn. 5 for $\Psi^+$ in this paper should be read as
$\sin{(|{\bf k}.{\bf r}|)/2}$; similarly $E_g(T)/Nk_BT$
in Eqn 25 should read as $E_g(T)/Nk_B$.  
\bibitem[43]{Kn:gnus}
Y.S. Jain, Ground state of $N$ hard core particles in 1-D box,
presented in CMDAYS-04 (Department of Physics, North-Eastern Hill
University, Shillong-793 022 (Meghalaya, India) : 
(A brief discussion included in [42]).
\bibitem[44]{Kn:gnus}
A. Bianconi, {\it et al} J. Phys.: Condens. Matter {\bf 12}, 10655
(2000) and references therein; N.L. Saini, H. Oyanagi, A. Lanzara,
D. Di Castro, S. Agrestini, A. Bianconi, F. Nakamura
and T. Fujita, Phys. Rev. Lett. {\bf 64} 132510 (2001).
\bibitem[45]{Kn:gnus}
A. Lanzara, G.M. Zhao, N.L. Saini, A. Bianconi, K. Conder,
H. Keller,
and K.A. Muller, J. Phys.: Condens. Matter {\bf 11}, L541
(1999); Z.-X. Shen, A. Lanzara, S. Ishihara and N. Nogaosa,
Phil. Mag. B {\bf 82}, 1349 (2002). 
\bibitem[46]{Kn:gnus}
(a) E.Z. Kuchinskii and M.V. Sadovskii, JETP {\bf 88}
968 (1999);         
(b) J. Schmalian, D.
Pines, and B. Stojkovic, Phys. Rev. B{\bf 60}, 667 (1999);
(c) V.M. Loktev, R.M. Quick and S.G. Sharapov,
Phys. Rep. {\bf 349}, 2 (2001);
(d) A. Perali, P. Pieri, G.C. Strinati and C. Castellani,
Phys. Rev. B{\bf 66}, 024510 (2002),  Y. Yanase, T. Jujo,
T. Nomura, H. Ikeda, T. Hotta and K. Yamada, Phys. Rep. {\bf 387},
1 (2004);
\bibitem[47]{Kn:gnus}
K. Huang, {\it Statistical Mechanics}, Wiley Eastern Ltd.
New Delhi (1991).
\bibitem[48]{Kn:gnus}
W.P. Halperin, Rev. Mod. Phys. {\bf 58}, 533 (1986).
\bibitem[49]{Kn:gnus}
G.E. Uhlenbeck and L. Gropper, Phys. Rev. {\bf 41}, 79 (1932);
the application of this method to a system of $N$ hard core
particles where each particle is represented by a macro-orbital
can be justified because a macro-orbital is nothing but the
superposition of two plane waves and the hard core interaction
is totally screened out in the state of such superposition.
The related derivations and analysis are reported in Ref.[50].  
\bibitem[50]{Kn:gnus}
Y.S. Jain, J. Sc. Exploration {\bf 16} 77 (2002).
\bibitem[51]{Kn:gnus}
R.K. Pathria, {\it statistical Physics}, Pergamon Press
Oxford (1977).  
\bibitem[52]{Kn:gnus}
J. Wilks, {\it The properties of Liquid and Solid Helium},
Clarendon
Press, Oxford (1967).
\bibitem[53]{Kn:gnus}
Y.S. Jain, {\it Wave mechanics of three different systems of
a particle in 1-D box and a 1-D oscillator} (Technical Report
SSP/YSJ-02 (2005).  From this report, we present the following
argument to follow the result in question.

Consider a system of :(i) a quantum particle of mass $m$ and
(ii) a 1-D oscillator (a particle of mass $M$ attached to a
spring of length $a$ and force constant $C$) placed side by side
in a 1-D box of length $l$ with both being in their ground
states;  they do not share any coordinate $x$ in the box with
the former representing a particle trapped in
a 1-D box of size $d = l-a$.  Evidently, we have
$E_o = h^2/8md^2 + 0.5\hbar\omega$ (with $\omega = \sqrt{C/M}$)
as ground state energy of the system. However, under the zero
point force of the particle, the length of the oscillator is
contracted by $\Delta{d}= d'-d$ and the effective size of the
box for quantum particle becomes $d'$.  Consequently, to a good
approximation, $E_o$ changes to $E'_o = h^2/8md'^2 +
0.5\hbar\omega + 0.5C\Delta{d}^2$ and use of the equilibrium
condition $f_o = h^2/4md'^3 = C\Delta{d}$, renders:
(i) $\Delta E = E'_o - E_o
= - (h^2/8md^3)(\Delta{d}/d)$ as the net decrease in the energy of
the system, (ii) $\Delta\varepsilon_o= - (h^2/4md^3)(\Delta{d})$
as the net decrease in the zero-point energy of the quantum
particle and (iii) $0.5C\Delta{d}^2 = + (h^2/8md^3)(\Delta{d})$ as
the net energy added to the spring as its strain energy.
\bibitem[54]{Kn:gnus}
J.C. wheatley in {\it The Helium Liquids} edited by J.G.M.
Armitage and I.E. Farquhar, Academic Press,
London (1975) pp 241-313.
\bibitem[55]{Kn:gnus}
C. Enss and S. Hunklinger, {\it Low Temperature Physics},
Springer-Verlag, Berlin 2005.
\bibitem[56]{Kn:gnus}
C.M. Varma, Z. Nussinov and W. van Saarloos, 
e-print : arXiv:cond-mat/0103393, H.J. Schulz, G. Cuniber
and P. Pieri e-print : arXiv:cond-mat/9807366, D. Vollhardt,
e-print : arXiv:cond-mat/9706100.
\bibitem[57]{Kn:gnus}
A.J. Leggett, Phys. Rev. Lett. {\bf 83}, 392 (1999).
\bibitem[58]{Kn:gnus}
A. Mourachkine, {\it Principles of Superconductivity},
e-print : arXiv:cond-mat/0409240.
\bibitem[59]{Kn:gnus}
L.D. Landau, J. Phys.(USSR) {\bf 5}, 71 (1941); english
translation published in {\it Helium 4} by Z.M. Galasiewicz,
(Pergamon Press, Oxford, 1971) pp 191-233.
\bibitem[60]{Kn:gnus}
J. Bardeen, Phys. Rev. Lett. {\bf 1}, 399 (1958). 

\end{thebibliography}
\end{document}